\documentclass[floatfix,prd,10pt,aps]{revtex4-2}
\usepackage{graphicx,epstopdf}
\usepackage[caption=false]{subfig}
\usepackage{appendix}
\usepackage[T1]{fontenc}
\usepackage{lmodern}
\usepackage[dvipsnames,x11names]{xcolor}
\usepackage[sort&compress]{natbib}
\usepackage{morefloats}
\usepackage{amsmath}
\usepackage{amssymb}
\usepackage{amsfonts}
\usepackage{mathtools}
\usepackage{bbold}
\usepackage{multirow}
\usepackage[normalem]{ulem}
\usepackage{physics}
\usepackage{colortbl}
\usepackage{placeins}
\usepackage{booktabs}
\usepackage{slashed}
\definecolor{linkc}{RGB}{0,70,140}
\definecolor{citec}{RGB}{170,80,0}
\usepackage[colorlinks=true,linkcolor=linkc,citecolor=citec,urlcolor=linkc]{hyperref}
\usepackage{orcidlink}
\allowdisplaybreaks   

\def\p_1slash{p1\!\!\!\slash }
\def\p_2slash{p2\!\!\!\slash }

\begin{document}

\title{Quark and gluon gravitational form factors of the nucleon at next-to-leading order}

\author{Ula\c{s} \"{O}zdem\orcidlink{0000-0002-1907-2894}}
\email{ulasozdem@aydin.edu.tr}
\affiliation{Health Services Vocational School of Higher Education, Istanbul Aydin University, Sefakoy-Kucukcekmece, 34295 Istanbul, T\"{u}rkiye}

\begin{abstract}
We compute the quark and gluon gravitational form factors $A$, $J$ and $D$ of the nucleon in three-point QCD sum rules at next-to-leading order in $\alpha_s$ for the quark and gluon energy-momentum tensors (EMT), treating both at the same order in the $\overline{\rm MS}$ scheme. The nucleon EMT also tests the method, showing where the sum rules are reliable and why. Protected estimators of $D$ and of the trace form factor $\bar c$ require antisymmetric Borel moments, since none with equal Borel masses alone is protected. The $\mathcal{O}(\alpha_s)$ corrections reduce $A_{\rm tot}$ by $32$--$40\%$, $J_{\rm tot}$ by $22$--$27\%$ and $|D_{\rm tot}|$ by $21$--$27\%$, since the perturbative share of the two-point function is much larger. The total form factors agree with lattice QCD for $1\le Q^2\le2.5$~GeV$^2$. At $Q^2=1.5$~GeV$^2$, $A_{\rm tot}=0.337\pm0.113$, $J_{\rm tot}=0.206\pm0.100$ and $D_{\rm tot}=-0.206\pm0.121$. $D$ is the most reliable result: its correlator is mostly perturbative, and its quark and gluon parts agree with the lattice. In $A$ and $J$, a local gluon-condensate term that does not fall with $Q^2$ is the largest, and their gluon shares are well below the lattice, possibly because the gluon EMT couples at this order only through gluon exchange and the condensate. A working region exists only near $\beta=1$, and the constant part of transitions cannot be separated within the Borel windows. The individual conservation identities deviate at one point by two standard deviations, consistent with a missing finite term in the $g_{\mu\nu}$ structure, and $\bar c$ is therefore given at leading order with condensates.
\end{abstract}

\maketitle

\section{Introduction}\label{sec:intro}

The matrix elements of the QCD energy-momentum tensor (EMT) between nucleon states are parametrized, for each constituent $a=q,g$, in terms of the gravitational form factors (GFFs) $A_a(t)$, $B_a(t)$, $D_a(t)$ and $\bar c_a(t)$~\cite{Kobzarev:1962wt,Pagels:1966zza}. They are given by the second Mellin moments of generalized parton distributions (GPDs)~\cite{Muller:1994ses,Ji:1996ek,Radyushkin:1996nd}, see Refs.~\cite{Goeke:2001tz,Diehl:2003ny} for reviews, and they enter the decomposition of the nucleon spin~\cite{Ji:1996ek,Leader:2013jra} and mass~\cite{Ji:1994av,Lorce:2017xzd,Metz:2020vxd} into quark and gluon contributions. For the total EMT, $A(0)=1$ and $B(0)=0$ follow from Poincaré invariance~\cite{Lowdon:2017idv}. The $D$-term~\cite{Polyakov:1999gs} is not fixed by any conservation law, and it has been related to the distribution of internal forces, pressure and shear in the nucleon and to its mechanical stability~\cite{Polyakov:2002yz,Polyakov:2018zvc}. The spatial distributions of energy, angular momentum and forces have been studied in various frames~\cite{Freese:2021czn,Lorce:2022cle}. Their definition and interpretation, including the notion of pressure, are still under discussion~\cite{Ji:2021mfb,Freese:2021qtb,Lorce:2025oot}, and a recent review can be found in Ref.~\cite{Burkert:2023wzr}. The trace form factor $\bar c_a$ is not constrained by forward sum rules either. It vanishes for the total EMT by conservation, it describes the force between the quark and gluon subsystems~\cite{Polyakov:2018exb}, and it is related to the trace anomaly and to the decomposition of the nucleon mass~\cite{Hatta:2018sqd,Lorce:2017xzd,Metz:2020vxd,Tanaka:2022wzy}.

Experimental information on the quark GFFs is obtained from deeply virtual Compton scattering (DVCS) measured at Jefferson Lab~\cite{CLAS:2015uuo,CLAS:2026lls}, see Ref.~\cite{Kumericki:2016ehc} for a review. The quark contribution to the $D$-term was first extracted from these data~\cite{Burkert:2018bqq}, and the analysis was later extended to the shear forces and the mechanical radius~\cite{Burkert:2021ith}, see also Refs.~\cite{Kumericki:2019ddg,Dutrieux:2021nlz}. Quark GFFs and mechanical properties have also been obtained from global analyses of GPDs~\cite{Hashamipour:2022noy,Goharipour:2024atx} and from other phenomenological studies~\cite{Selyugin:2023hqu}. Near-threshold $J/\psi$ photoproduction has been measured~\cite{GlueX:2023pev,Duran:2022xag}, and its relation to the gluon GFFs and to the mass structure of the proton has been investigated in a number of works~\cite{Kharzeev:2021qkd,Hatta:2018ina,Mamo:2022eui,Guo:2023pqw}. The role of the reaction mechanism in these processes has also been examined~\cite{Du:2020bqj,JointPhysicsAnalysisCenter:2023qgg}. In addition, the gluon GFFs at large momentum transfer have been computed in perturbative QCD~\cite{Tong:2021ctu}, and the strangeness $D$-term has been related to near-threshold $\phi$-meson production~\cite{Hatta:2021can}. Future facilities are expected to extend these measurements considerably~\cite{AbdulKhalek:2021gbh,Accardi:2023chb}. 
On the lattice, the quark contributions to the nucleon GFFs were computed first~\cite{Mathur:1999uf,Hagler:2003jd,LHPC:2007blg}. These calculations were followed by quark and gluon decompositions of the momentum and spin~\cite{Yang:2018bft,Alexandrou:2020sml}, see Ref.~\cite{Liu:2021lke} for a review, and by studies of the decomposition of the nucleon mass~\cite{Yang:2018nqn}, the gluon GFFs~\cite{Shanahan:2018pib,Pefkou:2021fni}, the pressure distribution~\cite{Shanahan:2018nnv} and trace-anomaly form factors~\cite{Wang:2024lrm}. Recently, a complete flavour decomposition into up, down, strange and gluon contributions has been obtained at a pion mass close to its physical value~\cite{Hackett:2023rif}. 
The GFFs have also been studied in a wide range of approaches, including chiral effective field theory~\cite{Belitsky:2002jp,Alharazin:2020yjv}, the chiral quark-soliton model~\cite{Goeke:2007fp,Won:2023zmf}, Skyrme and related soliton models~\cite{Cebulla:2007ei,Tanaka:2025pny}, the bag model~\cite{Neubelt:2019sou}, a large-$N_c$ quark model~\cite{Lorce:2022cle}, dispersion relations~\cite{Pasquini:2014vua,Cao:2024zlf}, holographic QCD~\cite{Abidin:2009hr,Mamo:2022eui}, light-front and quark--diquark models~\cite{Chakrabarti:2020kdc,Choudhary:2022den}, basis light-front quantization~\cite{Nair:2025sfr}, the Nambu--Jona-Lasinio model~\cite{Freese:2020mcx,Qiu:2026kde}, the Faddeev equation~\cite{Yao:2024ixu} and meson dominance~\cite{Broniowski:2025ctl}. The role of the scale anomaly in the internal pressure has been examined in Ref.~\cite{Tanaka:2025pny}. For charged particles, electromagnetic forces change the small-$t$ behaviour of $D(t)$~\cite{Varma:2020crx,Metz:2021lqv}. The GFFs of other hadrons and systems have been studied as well, for example those of the pion~\cite{Hackett:2023nkr}, of spin-$3/2$ baryons~\cite{Kim:2020lrs,Dehghan:2023ytx}, of nuclei~\cite{GarciaMartin-Caro:2023klo}, and of model systems such as Q-balls~\cite{Mai:2012yc}.

Within QCD sum rules~\cite{Shifman:1978bx,Ioffe:1981kw}, the quark GFFs of the nucleon have been obtained from light-cone sum rules with nucleon distribution amplitudes~\cite{Anikin:2019kwi,Azizi:2019ytx}, including a flavour decomposition~\cite{Dehghan:2025ncw}. The GFFs of other baryons~\cite{Ozdem:2020ieh,Azizi:2020jog,Dehghan:2023ytx,Dehghan:2025eov} and the gravitational transition form factors between the nucleon and its resonances~\cite{Ozdem:2019pkg,Ozdem:2022zig} have also been studied with QCD sum rules. In Ref.~\cite{Asmaee:2026nhr}, the gluon GFFs have been analyzed with three-point sum rules, where the gluon EMT is coupled to the quark lines through the gluon condensate. To our knowledge, these sum-rule analyses do not include the perturbative $\mathcal{O}(\alpha_s)$ corrections. 
In this study, we apply the three-point QCD sum-rule method~\cite{Ioffe:1982qb,Nesterenko:1982gc} to the nucleon GFFs at next-to-leading order in $\alpha_s$, for both the quark and the gluon EMT. At this order the gluon EMT couples perturbatively to the quark lines through the exchanged gluon. As a result, both constituents are treated at the same order in the $\overline{\rm MS}$ scheme. The analysis is based on three methodological elements. First, the form factors are extracted from combinations of Lorentz structures that are exact for the double pole and are protected by the Ward identities in the forward limit (Sec.~\ref{sec:formalism}). Second, we show that no estimator of $\bar c$ based on equal Borel masses alone can be protected by EMT conservation. Instead, we construct protected estimators of $\bar c$ and $D$ from the conservation identities in Borel space, which require antisymmetric Borel moments, and we compute these moments at $\mathcal{O}(\alpha_s)$ (Sec.~\ref{sec:conservation}). Third, we check the $\mathcal{O}(\alpha_s)$ three-point functions with a set of internal consistency tests that follow from EMT conservation and renormalization-group invariance (Sec.~\ref{sec:validation}). We compare the obtained results with lattice QCD and other approaches in Sec.~\ref{sec:compare}.

\section{Formalism}\label{sec:formalism}

\subsection{Energy-momentum tensor and form factors}

The quark and gluon parts of the (symmetric, gauge-invariant) EMT are
\begin{equation}
T^q_{\mu\nu}=\frac{i}{4}\,\bar q\big(\gamma_\mu\overleftrightarrow{D}_\nu+\gamma_\nu\overleftrightarrow{D}_\mu\big)q,\qquad
T^g_{\mu\nu}=-G^a_{\mu\alpha}G^{a\,\alpha}_{\nu}+\frac14\,g_{\mu\nu}\,G^a_{\alpha\beta}G^{a\,\alpha\beta},
\end{equation}
and their nucleon matrix elements are decomposed as
\begin{equation}
\begin{split}
\langle p'|T^a_{\mu\nu}|p\rangle=\bar u(p')\Big[&A_a\,\gamma_{(\mu}P_{\nu)}+B_a\,\frac{P_{(\mu}i\sigma_{\nu)\alpha}q^\alpha}{2m}\\
&+D_a\,\frac{q_\mu q_\nu-g_{\mu\nu}q^2}{4m}+\bar c_a\, m\, g_{\mu\nu}\Big]u(p),
\end{split}
\label{eq:decomp}
\end{equation}
with $P=(p+p')/2$, $q=p'-p$, $t=q^2=-Q^2$, $X_{(\mu}Y_{\nu)}=\tfrac12(X_\mu Y_\nu+X_\nu Y_\mu)$ and $J_a=(A_a+B_a)/2$. Conservation of the total EMT implies $\sum_a\bar c_a(t)=0$ for all $t$, together with $A_{\rm tot}(0)=1$ and $J_{\rm tot}(0)=1/2$.

\subsection{Correlation function}

For the nucleon we use the general interpolating current without derivatives,
\begin{equation}
\eta(x)=2\epsilon^{abc}\Big[\big(u^{aT}C d^b\big)\gamma_5 u^c+\beta\big(u^{aT}C\gamma_5 d^b\big)u^c\Big],
\label{eq:current}
\end{equation}
where $\beta=-1$ gives the Ioffe current. The correlation function is
\begin{equation}
\Pi_{\mu\nu}(p,p')=i^2\int d^4x\,d^4y\;e^{ip'\cdot x-ip\cdot y}\,\langle0|\mathcal{T}\{\eta(x)\,T_{\mu\nu}(0)\,\bar\eta(y)\}|0\rangle ,
\end{equation}
which we expand in the structures with an odd number of Dirac matrices,
\begin{equation}
\begin{aligned}
&S_1=p_\mu p_\nu\slashed{p},\quad S_2=p_\mu p_\nu\slashed{p}',\quad S_5=p_\mu p'_\nu\slashed{p},\quad S_7=g_{\mu\nu}\slashed{p},\\
&S_9=\gamma_\mu p_\nu,\quad S_{11}=\gamma_{[p}\gamma_{p'}\gamma_{\mu]}\,p_\nu ,
\end{aligned}
\label{eq:structures}
\end{equation}
together with their $p\leftrightarrow p'$ partners ($S_3$, $S_4$, $S_6$, $S_8$, $S_{10}$, $S_{12}$) and the $\mu\leftrightarrow\nu$ images. For equal Borel masses the $p\leftrightarrow p'$ symmetry gives $C_3=C_2$, $C_4=C_1$, $C_6=C_5$, $C_8=C_7$, $C_{10}=C_9$. Table~\ref{tab:pole} lists the residues of the double pole in the decomposition~\eqref{eq:decomp}.

\begin{table}[htb]
\caption{Coefficients of the double-pole residue in the structures~\eqref{eq:structures} (normalization $\lambda^2e^{-2m^2/M^2}$ omitted). The coefficients of $S_9$ and $S_{10}$ follow from conservation of the pole term, $C_9=-Q^2J/4$.}
\label{tab:pole}
\renewcommand{\arraystretch}{1.3}
\begin{ruledtabular}
\begin{tabular}{lcccc}
Structure & $A$ & $B$ & $D$ & $\bar c$\\
\hline
$S_1$, $S_4$ & $0$ & $-1/4$ & $1/4$ & $0$\\
$S_2$, $S_3$ & $1/2$ & $1/4$ & $1/4$ & $0$\\
$S_5$, $S_6$ & $1/4$ & $0$ & $-1/4$ & $0$\\
$S_7$, $S_8$ & $0$ & $0$ & $Q^2/4$ & $m^2$\\
$S_{11}$, $S_{12}$ & $1/4$ & $1/4$ & $0$ & $0$\\
\end{tabular}
\end{ruledtabular}
\end{table}

We apply a double Borel transformation in $p^2$ and $p'^2$ with equal Borel masses, $M_1^2=M_2^2=M^2$ ($\tau=1/M^2$), and normalize the hadronic side as $H_i=\Pi_i\,e^{2m^2/M^2}/\lambda^2$, with $\lambda^2$ taken from the two-point sum rule at the same order.

\subsection{Ward-protected combinations}

The combinations $A=2(H_{11}+H_1+H_5)$ and $J=2H_{11}$ are exact for the pole but do not suit the operator product expansion (OPE). Already in the leading-order (LO) perturbative OPE, the coefficients $C_1$ and $C_5$ each contain parts that grow like $1/Q^2$ at small $Q^2$ and nearly cancel. Since the quark EMT is the total EMT at LO, $A(0)=1$ and $J(0)=1/2$ must hold there, and only the combinations
\begin{equation}
A=H_1+H_2+2H_5,\qquad J=2H_{11},\qquad B=2J-A
\label{eq:protected}
\end{equation}
reproduce these values. They are exact for the pole (Table~\ref{tab:pole}) and tied to the conservation of the EMT in the forward limit, and we use them throughout.

\subsection{Conservation identities and the estimators of \texorpdfstring{$D$}{D} and \texorpdfstring{$\bar c$}{c-bar}}\label{sec:conservation}

The pole coefficients give $C_7=m^2\bar c+Q^2D/4$. The estimator $m^2\bar c_W=H_7-(Q^2/4)(H_1+H_2-2H_5)$, which uses equal Borel masses only, violates $\bar c_{\rm tot}=0$ even in exactly conserved correlators, and the conservation identities in Borel space show why. The components of $q^\mu\Pi_{\mu\nu}$ along $p_\nu\slashed p$ and $p'_\nu\slashed p$ give, with $B[p^2f]=-\partial_{\tau_1}B[f]$ and $B[p'^2f]=-\partial_{\tau_2}B[f]$ (contact terms vanish under the double Borel transformation),
\begin{align}
V_1&\equiv\tfrac12\mathcal{A}(B_1+B_5)+\tfrac{Q^2}{2}(B_1-B_5)-B_7-B_9=0,\label{eq:V1}\\
V_3&\equiv\tfrac12\mathcal{A}(B_3+B_5)+\tfrac{Q^2}{2}(B_5-B_3)+B_7-B_{10}=0,\label{eq:V3}
\end{align}
where $\mathcal{A}=\partial_{\tau_1}-\partial_{\tau_2}$ is evaluated at $\tau_1=\tau_2$. Since the double pole behaves as $e^{-m^2(\tau_1+\tau_2)}$, $\mathcal{A}$ annihilates it, and the pole gives $V_1=-m^2\bar c\,\Lambda$ and $V_3=m^2\bar c\,\Lambda$ with $\Lambda=\lambda^2e^{-m^2(\tau_1+\tau_2)}$. For a general conserved correlator the equal-mass estimator equals the antisymmetric derivative, and data at $\tau_1=\tau_2$ give no access to it. An estimator of $\bar c$ built from equal Borel masses alone is therefore not protected by conservation. The protected estimators are
\begin{equation}
m^2\bar c^{(1)}=-V_1\,n,\qquad m^2\bar c^{(3)}=V_3\,n,\qquad D^{(i)}=\frac{4}{Q^2}\big(H_7-m^2\bar c^{(i)}\big),
\label{eq:estimators}
\end{equation}
with $n=e^{2m^2/M^2}/\lambda^2$. They give $\bar c^{(i)}=\bar c$ and $D^{(i)}=D$ for the pole and $\bar c^{(i)}=0$ exactly for every conserved correlator, and for the total EMT the two estimators coincide.

For the present analysis it matters that the $D$ estimators of Eq.~\eqref{eq:estimators} do not depend on the $g_{\mu\nu}$ structure at all. Inserting $V_1$ and $V_3$ gives
\begin{equation}
\begin{aligned}
D^{(1)}_a&=\frac{4n}{Q^2}\Big[\tfrac12\mathcal{A}(B_1+B_5)+\tfrac{Q^2}{2}(B_1-B_5)-B_9\Big]_a,\\
D^{(3)}_a&=\frac{4n}{Q^2}\Big[-\tfrac12\mathcal{A}(B_3+B_5)-\tfrac{Q^2}{2}(B_5-B_3)+B_{10}\Big]_a,
\end{aligned}
\label{eq:Dnoseven}
\end{equation}
for each constituent $a$, since $B_7$ cancels between $H_7$ and $V_{1,3}$. The estimators of $\bar c$, in contrast, contain $B_7$ directly.

\section{Operator product expansion}\label{sec:ope}

\subsection{Leading order and condensates}

\begin{figure}[htb]
\centering
\includegraphics[width=0.95\textwidth]{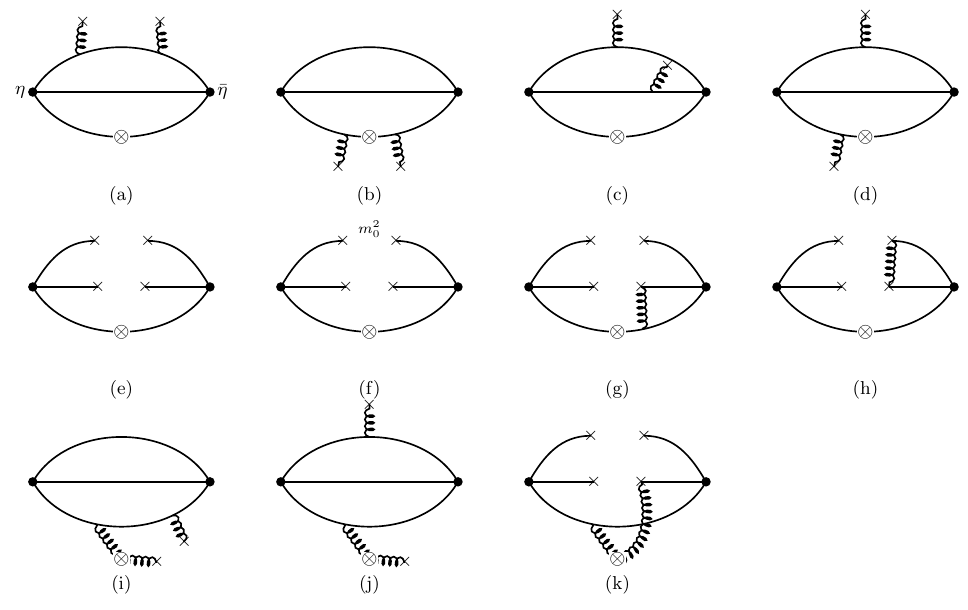}
\caption{LO condensate diagrams in the local OPE, in the Fock--Schwinger gauge centred at the EMT vertex. The symbol $\times$ denotes a vacuum field, and quark lines ending in $\times$ the quark condensate. Quark EMT ($\otimes$ on the lower, active line): gluon condensate $\langle g^2G^2\rangle$ with (a) both gluons on one spectator line, (b) one gluon on each side of the EMT vertex on the active line, (c) one gluon on each spectator line, and (d) one gluon on the active and one on a spectator line; (e) four-quark condensate $\langle\bar qq\rangle^2$, with the two spectator lines replaced by the condensate; mixed condensate $m_0^2\langle\bar qq\rangle^2$ from (f) the $m_0^2$ term of the nonlocal expansion of a condensate line, (g) the gluon of the quark--gluon condensate absorbed by the active line, and (h) a soft gluon between the two condensate lines, which arises from the phase factor of the condensate lines in this gauge. Gluon EMT ($\otimes$ on the gluon lines): the hard gluon is absorbed by a quark line, and the soft gluon pairs (i) with a soft gluon on the same line or (j) on another line, or (k) is absorbed by the quark--gluon condensate of a spectator line, the other spectator line being replaced by the quark condensate. In this gauge the gluon field vanishes at the EMT vertex, and two soft gluons on the same segment of the active line do not contribute. Diagrams (g), (h) and (k) contribute to the structures used here only in the $d$-quark part, and the four-quark and mixed-condensate contributions are proportional to $\beta-1$. Permutations over the spectator lines are not shown.}
\label{fig:conddiagrams}
\end{figure}

At LO, the quark-EMT correlator receives the perturbative contribution and the contributions of the gluon condensate $\langle g^2G^2\rangle$ (dimension four), the four-quark condensate (dimension six) and the mixed condensate $m_0^2\langle\bar qq\rangle^2$ (dimension eight) in the local OPE. Since the gluon EMT couples to the quark lines only through gluon exchange, its perturbative contribution starts at $\mathcal{O}(\alpha_s)$, and the LO gluon correlator consists of condensate terms only. We subtract the continuum with the triangle model $\sigma+\sigma'\le S_0$ in the double spectral representation. The LO and condensate expressions are given in Appendix~\ref{app:ope}, and the condensate diagrams are shown in Fig.~\ref{fig:conddiagrams}.

\subsection{\texorpdfstring{$\mathcal{O}(\alpha_s)$}{O(alpha s)} corrections}

\begin{figure}[htb]
\centering
\includegraphics[width=0.95\textwidth]{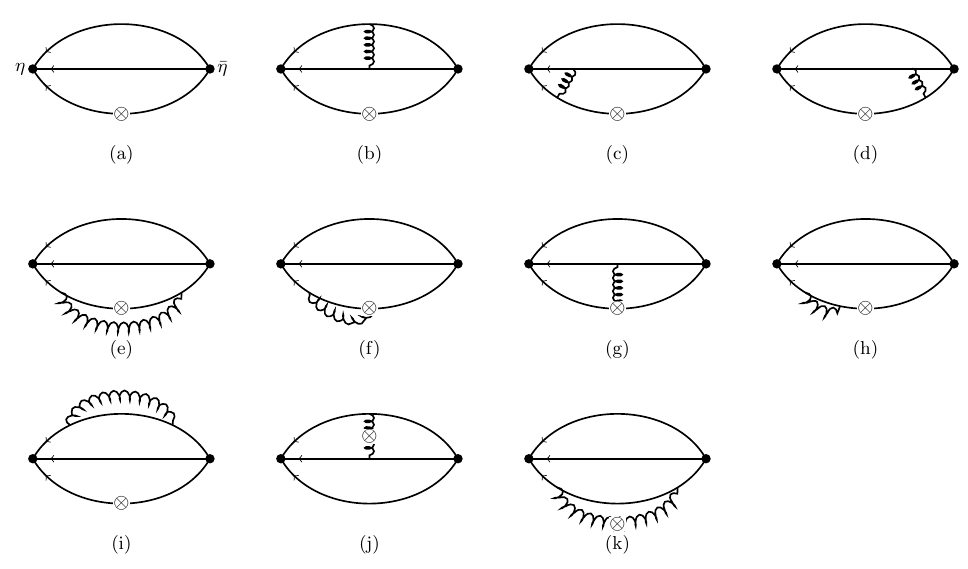}
\caption{LO and $\mathcal{O}(\alpha_s)$ diagrams of the three-point function. The lower quark line carries the quark EMT ($\otimes$); the dots denote the interpolating currents $\eta$ (left) and $\bar\eta$ (right). (a) LO. Quark EMT at $\mathcal{O}(\alpha_s)$: (b) gluon exchange between the two spectator lines; (c), (d) exchange between a spectator line and the active line on the $\eta$ and on the $\bar\eta$ side (two spectators each); (e) one-loop correction of the quark-EMT vertex; (f) contact diagram with the gluon from the EMT vertex attached to the active line (either side); (g) contact diagram with the gluon attached to a spectator line (two spectators); (h) self-energy of the active line (either segment); (i) self-energy of a spectator line (two spectators). Gluon EMT at $\mathcal{O}(\alpha_s)$: (j) gluon EMT inserted on a gluon exchanged between two different quark lines ($uu$ and $ud$ pairs); (k) gluon EMT inserted on a gluon emitted and reabsorbed by the same quark line ($u$ or $d$). Each quark-EMT topology occurs with the EMT on a $u$ line and on the $d$ line. These are all topologies at this order: the gluon EMT couples only to gluon lines, and its three-gluon vertex does not enter at $\mathcal{O}(\alpha_s)$.}
\label{fig:diagrams}
\end{figure}

At $\mathcal{O}(\alpha_s)$, the quark-EMT correlator receives contributions from gluon exchange between the quark lines of the currents, from the one-loop correction to the quark-EMT vertex, from contact diagrams in which the gluon is attached to the EMT vertex, and from the quark self-energies. The gluon-EMT correlator at this order is obtained by inserting the gluon EMT on the exchanged gluon. All topologies at this order are shown in Fig.~\ref{fig:diagrams}.

We renormalize the quark self-energies and the EMT in the standard $\overline{\rm MS}$ scheme, which fixes the meaning of the scale-dependent decomposition $A_q(\mu)$, $A_g(\mu)$. For the subgraphs of the interpolating currents, we use a structure-preserving $\overline{\rm MS}$ prescription. In this prescription the counterterm is the pole times the full $D$-dimensional local structure of the subgraph, without four-dimensional Fierz reduction, and $\gamma_5$ is removed from the current strings by anticommutation, consistent with naive dimensional regularization. This prescription differs from the standard one by a finite renormalization of the currents, which is common to the two- and three-point functions and cancels in the form factors. The forward Ward test of Sec.~\ref{sec:validation} supports this cancellation.

The divergent subgraphs are subtracted locally and added back analytically. For the remaining finite integrals, the Dirac algebra is evaluated in four dimensions. The scheme dependence of this choice resides entirely in the divergent subgraphs, and it is removed by adding back the finite parts of the $D$-dimensional local vertex functions. We have recomputed these local vertex functions with the Dirac algebra in $D$ dimensions and with the algebra fixed to four dimensions. In the structure $\tfrac12(q_\mu\gamma_\nu+q_\nu\gamma_\mu)$, the difference equals the finite constants used in the production, namely $-2/3$ for the quark-EMT vertex and $-8/3$ for the gluon-EMT vertex, while the structure $q_\mu q_\nu\slashed q/q^2$ receives no shift. Therefore, the $p_\mu p_\nu$ sector is in the standard $\overline{\rm MS}$ scheme.

We evaluate the finite integrals numerically at seven values of $x=Q^2\tau$, $x\in\{0.5,0.7,1.25,1.8,2.4,3.0,3.5\}$, at a reference Borel parameter $\tau_0$, and use the homogeneity of the correlator,
\begin{equation}
B(\tau,Q^2,S_0)=\lambda^{-d}\big[B(\tau_0,Q^2\lambda,S_0\lambda)+\ln\lambda\;C_\mu(\tau_0,Q^2\lambda,S_0\lambda)\big],\qquad \lambda=\tau/\tau_0,
\label{eq:scaling}
\end{equation}
where $C_\mu$ is the coefficient of $\ln\mu^2$ and $d$ the mass dimension of the structure. We interpolate the $\mathcal{O}(\alpha_s)$ correction as a ratio to the LO correlator, which is then multiplied by the exact LO correlator. The antisymmetric moments $\mathcal{A}B_i$ carry one additional unit of mass dimension, and their LO part crosses zero in parts of the kinematic region. Therefore, we interpolate them as a ratio to the symmetric LO correlator of the same structure, with an additional factor $\lambda^{-1}$. At LO both interpolations reproduce the exact correlator to $\mathcal{O}(10^{-4})$ (median) in the analysis region.

\section{Consistency tests of the \texorpdfstring{$\mathcal{O}(\alpha_s)$}{O(alpha s)} correlators}\label{sec:validation}

The $\mathcal{O}(\alpha_s)$ three-point functions are subject to a set of tests that follow from EMT conservation and renormalization-group invariance, and these tests do not use any external input.

In the forward limit, the three-point function of the total EMT is fixed by the two-point function, both for the combination of $A$ and for that of $J$. For this purpose, we extrapolate the ratio of the $\mathcal{O}(\alpha_s)$ to the LO three-point function, computed at $S_0=\infty$ at the seven kinematic points, to $x\to0$ with polynomials of second and third degree in $x$. For the total EMT, the obtained values are $21.55$--$21.84$ for $A$ and $21.27$--$21.48$ for $J$ (in units of $a=\alpha_sC_F/(4\pi)$, for $\beta=-1$, $0$ and $1$), compared with $21.51$ for the two-point function, which corresponds to an agreement within $1.5\%$. With the quark EMT alone, the ratios are $17.1$--$17.5$ and $16.8$--$17.3$, respectively, so that the identity is satisfied only once the gluon EMT is added.

The scale dependence of the $\mathcal{O}(\alpha_s)$ correlators is constrained by renormalization-group invariance. The sum of the $\ln\mu^2$ coefficients of all parts must equal the anomalous dimension of the currents times the LO correlator, $\sum C_\mu=3(\mathrm{LO}_u+\mathrm{LO}_d)$. In the analysis region, this holds to $\le2.5\times10^{-3}$ for the symmetric moments and to $\le1.4\times10^{-2}$ for the antisymmetric moments. Both are normalized to $|\mathrm{LO}_u|+|\mathrm{LO}_d|$, since the antisymmetric LO correlator nearly cancels between flavours at some points.

In addition to these tests, EMT conservation provides a further one. The combination $V_1+V_3$, in which $B_7$ cancels, must vanish for the total EMT. At all seven kinematic points and at all 21 values of $(S_0,\beta)$ considered, it is consistent with zero within the numerical uncertainty, $|V_1+V_3|/\sigma\le0.34$, while the quark and gluon parts are individually nonzero and have opposite signs in $145$ of the $147$ cases (Table~\ref{tab:conservation}).

Since the two estimators $D^{(1)}$ and $D^{(3)}$ in Eq.~\eqref{eq:Dnoseven} are built from different structures, their agreement is a test as well. For the total EMT, they agree to $\le0.006$ at all $Q^2$. We have also verified numerically that $D$ does not depend on $B_7$: replacing the $\mathcal{O}(\alpha_s)$ part of $B_7$ by an arbitrary multiple of its LO value leaves $D$ unchanged to $4\times10^{-17}$.

The individual identities $V_1=0$ and $V_3=0$ contain $B_7$, and therefore they test the $g_{\mu\nu}$ sector, which carries the trace form factor $\bar c$. This sector is the hardest one to control at $\mathcal{O}(\alpha_s)$. The trace parts of the quark and gluon tensors mix under renormalization, and their separation, which defines $\bar c_q$ and $\bar c_g$, is related to the decomposition of the trace anomaly and depends on the scheme~\cite{Hatta:2018sqd,Tanaka:2022wzy}. Only their sum, which vanishes, is scheme independent. We have computed the $\mathcal{O}(\alpha_s)$ coefficient of $S_7$ at one kinematic point, $x=1.25$, and at this point $V_1$ and $V_3$ deviate from zero by amounts of opposite sign equal to $3.2$--$3.4$ times the LO $S_7$ coefficient in the working region at $\beta=1$, with a numerical uncertainty of $1.6$--$1.9$ times that coefficient, i.e.\ at the level of two standard deviations ($2.1$--$2.4\sigma$ at $\beta=-1$). The uncertainty comes mainly from the terms of $V_1$ and $V_3$ other than $B_7$, in particular from the antisymmetric moments, and less from $S_7$ itself. For this reason, we have recomputed the diagram classes that dominate it with Monte Carlo budgets increased by factors of $4$--$16$. All of them agree with their previous values within $0.53$ of the previous error estimates. As a result, the uncertainty of the deviation decreased by a factor of about two, while the deviation itself changed by only $0.4$ times the LO coefficient, well within its previous uncertainty, and did not approach zero. Its ratio to the LO $S_7$ coefficient lies between $3.2$ and $3.4$ for all three values of $\beta$ and all thresholds $S_0$. This would be consistent with a missing finite term proportional to the LO structure, such as a finite renormalization of the trace part. We have excluded two candidate mechanisms: the trace-type component $g_{\mu\nu}\slashed q$ of the local vertex functions, which gives a contact term that vanishes under the double Borel transformation, and the product of the vertex pole with the $\mathcal{O}(\epsilon)$ part of the outer Dirac algebra. However, we have not identified the origin of the term. Whether the deviation is a constant multiple of the LO coefficient could be tested at a second kinematic point. At the other points, however, the numerical uncertainty of the antisymmetric moments is two to three times larger in units of the LO $S_7$ coefficient. Because of Eq.~\eqref{eq:Dnoseven}, none of this affects $D$, and the combinations for $A$, $J$ and $B$ do not contain $S_7$. On the other hand, it prevents an $\mathcal{O}(\alpha_s)$ determination of $\bar c$, which is therefore given at LO with condensates.

\begin{table}[htb]
\caption{Conservation tests of the $\mathcal{O}(\alpha_s)$ correlators at the seven kinematic points ($21$ values of $(S_0,\beta)$ each).}
\label{tab:conservation}
\renewcommand{\arraystretch}{1.3}
\small
\begin{ruledtabular}
\begin{tabular}{cccc}
$x$ & $\max|V_1+V_3|_{\rm tot}/\sigma$ & opposite-sign $q$/$g$ & $\mu$ rule (antisymmetric)\\
\hline
0.5 & 0.22 & 21/21 & $4.6\times10^{-4}$\\
0.7 & 0.34 & 19/21 & $1.2\times10^{-3}$\\
1.25 & 0.27 & 21/21 & $2.8\times10^{-3}$\\
1.8 & 0.31 & 21/21 & $6.8\times10^{-3}$\\
2.4 & 0.29 & 21/21 & $1.3\times10^{-2}$\\
3.0 & 0.29 & 21/21 & $1.4\times10^{-2}$\\
3.5 & 0.25 & 21/21 & $1.3\times10^{-2}$\\
\end{tabular}
\end{ruledtabular}
\end{table}

\section{Numerical analysis}\label{sec:numerics}

\begin{table}[htb]
\caption{Input parameters used in the numerical analysis, at the scale $\mu=1$~GeV. The light-quark masses are set to zero. The variations in the third column are those used in the uncertainty estimate.}
\label{tab:inputs}
\renewcommand{\arraystretch}{1.3}
\begin{ruledtabular}
\begin{tabular}{lccc}
Parameter & Value & Variation & Ref. \\
\hline
$m_N$                                  & $0.938$~GeV                   & --                                 & \cite{ParticleDataGroup:2024cfk} \\
$\alpha_s(1~\text{GeV})$               & $0.47$                        & scale variation                    & \cite{ParticleDataGroup:2024cfk} \\
$\langle\bar qq\rangle$                & $-(0.24)^3$~GeV$^3$           & $-(0.23)^3$, $-(0.25)^3$~GeV$^3$   & \cite{Belyaev:1982sa,Ioffe:2005ym} \\
$m_0^2$                                & $0.8$~GeV$^2$                 & $0.6$~GeV$^2$                      & \cite{Belyaev:1982sa,Ioffe:2005ym} \\
$\langle\alpha_sG^2/\pi\rangle$        & $0.012$~GeV$^4$               & $\times0.5$, $\times1.5$           & \cite{Shifman:1978bx,Ioffe:2005ym} \\
\end{tabular}
\end{ruledtabular}
\end{table}

The input parameters are listed in Table~\ref{tab:inputs}, where the value of $\alpha_s$ corresponds to its two-loop running from the world average~\cite{ParticleDataGroup:2024cfk}. We obtain the residue $\lambda^2$ from the two-point sum rule in the $\slashed q$ structure at the same order, with thresholds $s_0\in\{2.1,2.225,2.35\}$~GeV$^2$, a pole contribution of at least $50\%$ and a contribution of the highest-dimension condensate of at most $10\%$. The three-point Borel window is kept inside the two-point window for the same $\beta$ and $s_0$, so that the Borel variable of each channel stays within the window in which the two-point sum rule satisfies these criteria. With the two-point sum rule of Eq.~\eqref{eq:twopoint} and the parameters of Table~\ref{tab:inputs}, the residue is obtained as $\lambda^2=(9.6\pm1.8)\times10^{-4}$~GeV$^6$ at $\mathcal{O}(\alpha_s)$, i.e.\ $\lambda=0.031\pm0.003$~GeV$^3$, and as $\lambda^2=(5.3\pm1.2)\times10^{-4}$~GeV$^6$ at LO with condensates. The central value is the midpoint over $\beta\in\{0.95,1,1.05\}$ and the three values of $s_0$, and the uncertainty combines the half-range with the variation of the gluon condensate. The quark-condensate contributions to the two-point function are proportional to $\beta-1$, and for $\beta=0.95$ and $1.05$ their variation changes $\lambda^2$ by at most $3\%$, which is negligible here. For $\beta=1$ and $s_0=2.225$~GeV$^2$, the two-point window is $0.73\le M^2\le0.89$~GeV$^2$. Since the residue depends on $\beta$ and on the normalization of the current~\eqref{eq:current}, it cannot be compared directly with values obtained for the Ioffe current.

We determine the working region once, at the reference value $Q^2_{\rm ref}=1.5$~GeV$^2$, and apply it to all $Q^2$. The criteria, applied to the correlator of each form-factor combination, are a pole contribution $0.5\le\Pi(S_0)/\Pi(\infty)\le1$ and a four-quark fraction of at most $30\%$, with $S_0\in[3.0,4.7]$~GeV$^2$ for the triangle continuum. For $A$, the criteria remain satisfied in the reference cells at the other values of $Q^2$, in at least $99.5\%$ of the cells. For $J$, the four-quark criterion of the $u$ quark holds in $76\%$ and $69\%$ of the cells at $Q^2=2$ and $2.5$~GeV$^2$, respectively, and the pole criterion of the $d$ quark holds in $93$--$99\%$ of the cells at $Q^2\neq Q^2_{\rm ref}$. We keep these cells, since the working region is fixed at $Q^2_{\rm ref}$. The mixing parameter $\beta$ is scanned over $[-5,5]$ with a step of $0.05$. When a single interpolating current is required for all constituents, the criteria leave $\beta\in\{0.95,1,1.05\}$, and $A$ and $B$ are taken at $\beta=1$, which is the only value at which the criteria are satisfied for the combination of $A$.

The restriction to $\beta\approx1$ can be understood from the two-point sum rule. Its dimension-six and dimension-eight contributions are proportional to $(\beta-1)(7\beta+5)$ and $(\beta-1)(13\beta+11)$ [Eq.~\eqref{eq:twopoint}], respectively. At $s_0=2.225$~GeV$^2$, the two-point sum rule has a window only where the dimension-eight contribution, which is the highest one, is small compared with the perturbative contribution. This happens near the zeros of its coefficient at $\beta=1$ and $\beta=-11/13$, namely for $-1.10\le\beta\le-0.75$ and $0.85\le\beta\le1.30$. In the first interval, which contains the Ioffe current, the criteria for $J$ fail for the $d$ quark and for the gluon EMT. In the structure $S_{11}$, the perturbative coefficients of the $d$ quark are proportional to $(1+\beta)^2$ (Appendix~\ref{app:ope}) and vanish at $\beta=-1$. As a result, the $d$-quark correlator consists of condensate contributions only, and the gluon-EMT correlator is dominated by its dimension-eight part. In the second interval, the four-quark and mixed-condensate contributions are proportional to $\beta-1$, and the criteria are satisfied for all constituents only for $|\beta-1|\le0.05$. The working region thus lies where these contributions are suppressed by the structure of the current, and the convergence criterion is met largely for this reason. The dimension-four gluon-condensate contribution, on the other hand, is not suppressed (Table~\ref{tab:opecomp}). Therefore, the dependence of the results on the interpolating current can be assessed only within $\beta\in\{0.95,1,1.05\}$, where it is included in the spread over the working region. This dependence is sizeable, and the half-range of the central values of $J_{\rm tot}$ and $D_{\rm tot}$ over the three values of $\beta$ makes up $54$--$72\%$ of their half-range over the working region. At $Q^2=1.5$~GeV$^2$, for example, $J_{\rm tot}$ changes from $0.244$ to $0.126$ and $D_{\rm tot}$ from $-0.147$ to $-0.285$ between $\beta=0.95$ and $1.05$. This variation arises almost entirely from the four-quark and mixed-condensate contributions, which change sign at $\beta=1$. When these contributions are removed, it is reduced at $Q^2=1$ and $2.5$~GeV$^2$ to at most $0.013$ for $J_{\rm tot}$ and $0.010$ for $D_{\rm tot}$.

We obtain the uncertainty of each form factor as the quadratic sum of the following contributions: the spread over the working region ($\beta$, $s_0$, $S_0$ and $M^2$), the variation of $\langle g^2G^2\rangle$ by factors $0.5$ and $1.5$, the variation of the quark condensates ($\langle\bar qq\rangle=-(0.23)^3$ and $-(0.25)^3$~GeV$^3$, $m_0^2=0.6$~GeV$^2$), four numerical contributions from the evaluation of the $\mathcal{O}(\alpha_s)$ integrals (statistical error, reproducibility, convergence with the integration budget and interpolation, see Appendix~\ref{app:numerics}), and the scale variation $\mu^2\in\{0.5,1,2\}$~GeV$^2$ with two-loop running of $\alpha_s$. In the scale variation, the quark and gluon parts are evolved back to $\mu=1$~GeV with the first-order DGLAP evolution of the $n=2$ moments. For $D$ and $\bar c$, the central value is the average of the two estimators, and half of their difference is included as an additional systematic uncertainty. The transitions between the nucleon and excited states are discussed in Sec.~\ref{sec:limitations}.

\section{Results}\label{sec:results}

\begin{figure}[t]
\centering
\includegraphics[width=0.95\textwidth]{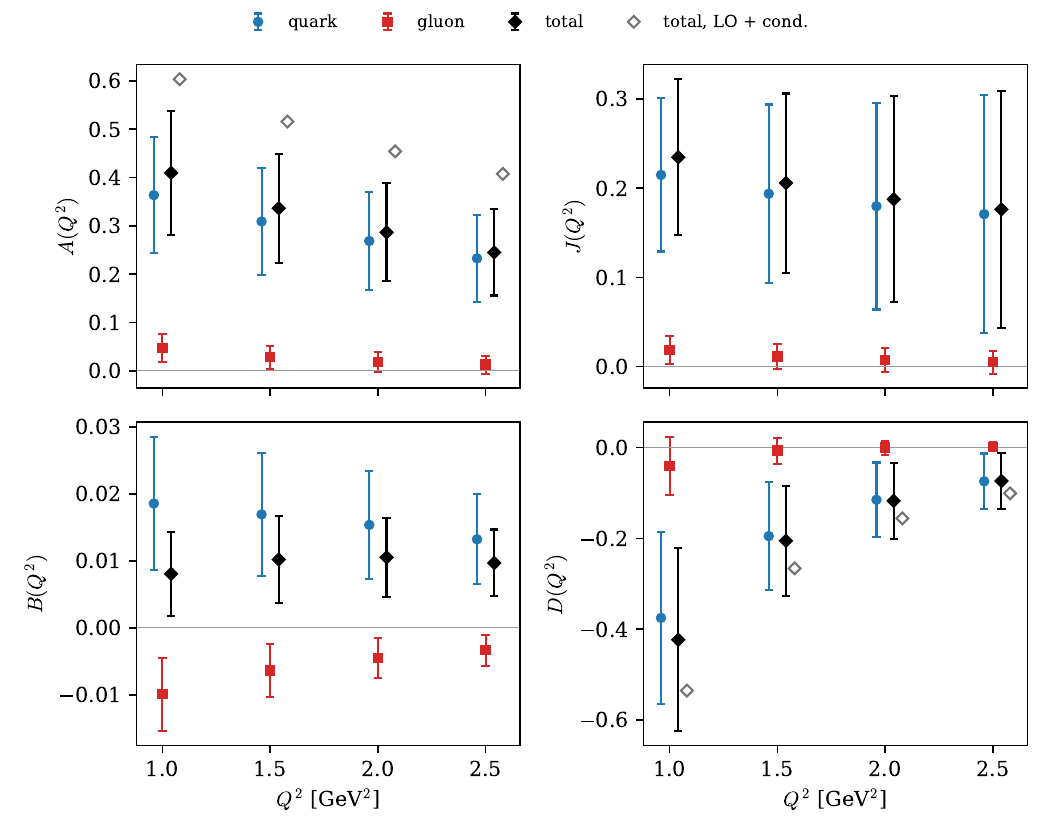}
\caption{Quark, gluon and total gravitational form factors at $\mathcal{O}(\alpha_s)$, $\mu=1$~GeV, $\overline{\rm MS}$. Open symbols: total $A$ and $D$ at LO with condensates. Points are displaced horizontally for visibility.}
\label{fig:gff}
\end{figure}

The obtained results at $\mathcal{O}(\alpha_s)$ for $A$, $J$ and $B$ are given in Table~\ref{tab:AJB}, and those for $D$ in Table~\ref{tab:D}. Their $Q^2$ dependence is shown in Fig.~\ref{fig:gff}, together with the total $A$ and $D$ of the LO analysis with condensates.

\begin{table}[htb]
\caption{Gravitational form factors $A$, $J$ and $B$ at $\mathcal{O}(\alpha_s)$, $\mu=1$~GeV, $\overline{\rm MS}$ ($Q^2$ in GeV$^2$). The uncertainty is the quadratic sum of the working-region spread, the condensate variations, the numerical uncertainty and the scale variation. Transitions are discussed in Sec.~\ref{sec:limitations}.}
\label{tab:AJB}
\renewcommand{\arraystretch}{1.3}
\footnotesize\setlength{\tabcolsep}{4pt}
\begin{ruledtabular}
\begin{tabular}{cccccc}
$Q^2$ & $A_u$ & $A_d$ & $A_q$ & $A_g$ & $A_{\rm tot}$\\
\hline
1.0 & $0.331\pm0.113$ & $0.032\pm0.011$ & $0.364\pm0.120$ & $0.047\pm0.029$ & $0.410\pm0.129$\\
1.5 & $0.286\pm0.103$ & $0.022\pm0.009$ & $0.309\pm0.110$ & $0.029\pm0.024$ & $0.337\pm0.113$\\
2.0 & $0.252\pm0.095$ & $0.016\pm0.007$ & $0.269\pm0.101$ & $0.019\pm0.021$ & $0.287\pm0.102$\\
2.5 & $0.220\pm0.085$ & $0.012\pm0.006$ & $0.233\pm0.090$ & $0.013\pm0.018$ & $0.245\pm0.089$\\
\hline
$Q^2$ & $J_u$ & $J_d$ & $J_q$ & $J_g$ & $J_{\rm tot}$\\
\hline
1.0 & $0.215\pm0.095$ & $-0.010\pm0.010$ & $0.215\pm0.086$ & $0.019\pm0.016$ & $0.235\pm0.088$\\
1.5 & $0.196\pm0.109$ & $-0.009\pm0.007$ & $0.194\pm0.100$ & $0.011\pm0.014$ & $0.206\pm0.100$\\
2.0 & $0.183\pm0.123$ & $-0.008\pm0.006$ & $0.180\pm0.116$ & $0.007\pm0.013$ & $0.188\pm0.116$\\
2.5 & $0.175\pm0.138$ & $-0.007\pm0.004$ & $0.171\pm0.133$ & $0.005\pm0.013$ & $0.176\pm0.133$\\
\hline
$Q^2$ & $B_u$ & $B_d$ & $B_q$ & $B_g$ & $B_{\rm tot}$\\
\hline
1.0 & $0.070\pm0.022$ & $-0.056\pm0.022$ & $0.019\pm0.010$ & $-0.010\pm0.005$ & $0.008\pm0.006$\\
1.5 & $0.057\pm0.018$ & $-0.044\pm0.016$ & $0.017\pm0.009$ & $-0.006\pm0.004$ & $0.010\pm0.006$\\
2.0 & $0.047\pm0.015$ & $-0.033\pm0.012$ & $0.015\pm0.008$ & $-0.005\pm0.003$ & $0.011\pm0.006$\\
2.5 & $0.039\pm0.013$ & $-0.026\pm0.010$ & $0.013\pm0.007$ & $-0.003\pm0.002$ & $0.010\pm0.005$\\
\end{tabular}
\end{ruledtabular}
\end{table}

\begin{table}[htb]
\caption{$D$ at $\mathcal{O}(\alpha_s)$, $\mu=1$~GeV, $\overline{\rm MS}$, from the estimators~\eqref{eq:Dnoseven}. The central value is the average of $D^{(1)}$ and $D^{(3)}$, and the uncertainty includes half of their difference.}
\label{tab:D}
\renewcommand{\arraystretch}{1.3}
\footnotesize\setlength{\tabcolsep}{3pt}
\begin{ruledtabular}
\begin{tabular}{cccccc}
$Q^2$ & $D_{\rm tot}$ & $D_q$ & $D_g$ & $D_u$ & $D_d$\\
\hline
1.0 & $-0.423\pm0.201$ & $-0.375\pm0.189$ & $-0.041\pm0.065$ & $-0.207\pm0.160$ & $-0.177\pm0.056$\\
1.5 & $-0.206\pm0.121$ & $-0.195\pm0.118$ & $-0.007\pm0.028$ & $-0.103\pm0.100$ & $-0.096\pm0.032$\\
2.0 & $-0.117\pm0.083$ & $-0.115\pm0.082$ & $0.000\pm0.016$ & $-0.059\pm0.070$ & $-0.058\pm0.020$\\
2.5 & $-0.074\pm0.062$ & $-0.075\pm0.061$ & $0.002\pm0.010$ & $-0.036\pm0.055$ & $-0.039\pm0.014$\\
\end{tabular}
\end{ruledtabular}
\end{table}

Compared with the LO analysis with condensates, the $\mathcal{O}(\alpha_s)$ corrections reduce $A_{\rm tot}$ by factors of $0.60$--$0.68$ and $J_{\rm tot}$ by factors of $0.73$--$0.78$. The relative perturbative correction is the same in the two- and three-point functions ($a\,O_1/\mathrm{LO}_{\rm pert}\approx1.09$--$1.15$ at $Q^2=1.5$~GeV$^2$), as required by the forward Ward identity. However, the perturbative part is about $75\%$ of the two-point function in its window, and only $17$--$24\%$ of the three-point $A_q$ correlator in the working region. As a result, the two-point function increases by a factor of $1.83$ and the three-point function by $1.20$--$1.26$. At $Q^2=1.5$~GeV$^2$, adding the $\mathcal{O}(\alpha_s)$ term to the two-point function alone gives $A_q=0.250$, to the three-point function alone $0.587$, and to both $0.309$, compared with $0.481$ at LO with condensates.

For $D$, the two estimators agree for the total EMT, as required by conservation. $D_{\rm tot}$ is almost independent of the scale ($\sigma_\mu\le0.004$) and is scheme independent, and the $\mathcal{O}(\alpha_s)$ correction reduces $|D_{\rm tot}|$ by $21$--$27\%$ relative to the LO analysis with condensates. $D_g$ is compatible with zero at all $Q^2$, and its uncertainty is dominated by the estimator systematic and by the scale variation. $D_u$ and $D_d$ are close to each other.

To understand which contributions determine the $Q^2$ dependence of these results, the composition of the $\mathcal{O}(\alpha_s)$ correlators in the working region is given in Table~\ref{tab:opecomp}. At $\beta=1$, the four-quark and mixed condensates do not contribute, since they are proportional to $\beta-1$. The correlators then consist of the perturbative part, at LO and at $\mathcal{O}(\alpha_s)$, and of the gluon-condensate contributions to the quark-EMT and gluon-EMT correlators. For $D$, the perturbative contribution is the largest one up to $Q^2=2$~GeV$^2$, and it is equal to the gluon-condensate one at $Q^2=2.5$~GeV$^2$. For $A$ and $J$, the gluon-condensate contribution to the quark-EMT correlator is the largest one, and its share grows from about one half to about two thirds over the range. In addition, it depends much more weakly on $Q^2$ than the perturbative part. At LO with condensates, its normalized contribution to $A$ and $J$ decreases by about $15\%$ between $Q^2=1$ and $2.5$~GeV$^2$, and by about $35\%$ up to $Q^2=5$~GeV$^2$, while the perturbative one decreases by factors of about $2.5$ and $6$--$7$, respectively. For $D$, the gluon-condensate and perturbative contributions decrease by factors of $3.5$ and $6.7$ between $Q^2=1$ and $2.5$~GeV$^2$. It is known that local-condensate contributions to three-point sum rules need not decrease with $Q^2$ like the perturbative part. In the sum rule for the pion form factor, the gluon-condensate contribution is constant and the quark-condensate one even grows with $Q^2$, while the perturbative term falls~\cite{Bakulev:1991ps,Bakulev:2009ib}, and this was one of the motivations for nonlocal condensates. Therefore, the $Q^2$ dependence of $A$ and $J$ in our range is to a large extent that of a condensate term without the fall-off expected for a form factor, whereas that of $D$ is governed by contributions that do show this fall-off. This limits the information that our results carry on the $Q^2$ dependence of $A$ and $J$ and on their forward values.

\begin{table}[htb]
\caption{Composition of the $\mathcal{O}(\alpha_s)$ correlators of $A_{\rm tot}$, $J_{\rm tot}$ and $D_{\rm tot}$ in the working region at $\beta=1$: shares of the perturbative contribution (LO and $\mathcal{O}(\alpha_s)$, quark and gluon EMT), of the gluon condensate in the quark-EMT correlator ($d_4$), and of the gluon condensate in the gluon-EMT correlator ($g_4$). The four-quark and mixed condensates do not contribute at $\beta=1$. The entries are medians over the cells and need not add up to one.}
\label{tab:opecomp}
\renewcommand{\arraystretch}{1.3}
\begin{ruledtabular}
\begin{tabular}{cccccccccc}
 & \multicolumn{3}{c}{$A_{\rm tot}$} & \multicolumn{3}{c}{$J_{\rm tot}$} & \multicolumn{3}{c}{$D_{\rm tot}$}\\
$Q^2$ & pert. & $d_4$ & $g_4$ & pert. & $d_4$ & $g_4$ & pert. & $d_4$ & $g_4$\\
\hline
1 & $0.45$ & $0.47$ & $0.08$ & $0.46$ & $0.49$ & $0.05$ & $0.66$ & $0.36$ & $-0.02$\\
1.5 & $0.40$ & $0.55$ & $0.06$ & $0.39$ & $0.57$ & $0.04$ & $0.62$ & $0.44$ & $-0.06$\\
2 & $0.34$ & $0.61$ & $0.04$ & $0.35$ & $0.63$ & $0.03$ & $0.58$ & $0.51$ & $-0.09$\\
2.5 & $0.31$ & $0.65$ & $0.04$ & $0.31$ & $0.67$ & $0.02$ & $0.55$ & $0.55$ & $-0.10$\\
\end{tabular}
\end{ruledtabular}
\end{table}

For convenience, we give representations of the central values in Table~\ref{tab:param}, with $n=2$ for $A$, $J$ and $B$ and $n=3$ for $D$ and $\bar c_q$. The uncertainties at each $Q^2$ are those of Tables~\ref{tab:AJB}, \ref{tab:D} and~\ref{tab:cbar}. The representations are normalized at $Q^2=1$~GeV$^2$, and they are not meant for extrapolation. Fits of the individual cells of the working region for $A_{\rm tot}$ and $D_{\rm tot}$ show that the forward value of $D$ is not constrained by our results, and that of $A$ only weakly.

\begin{table}[htb]
\caption{Representation of the central values in $1\le Q^2\le2.5$~GeV$^2$ by $F(Q^2)=F_1\,[(1+1/\Lambda^2)/(1+Q^2/\Lambda^2)]^n$, with $Q^2$ and $\Lambda^2$ in GeV$^2$. The representation reproduces the central values to within $5\%$ of their uncertainties and is not meant for extrapolation outside this range.}
\label{tab:param}
\renewcommand{\arraystretch}{1.3}
\begin{ruledtabular}
\begin{tabular}{lccc}
 & $n$ & $F_1$ & $\Lambda$ (GeV)\\
\hline
$A_{\rm tot}$ & 2 & $0.4083$ & $2.036$\\
$A_q$ & 2 & $0.3635$ & $2.248$\\
$A_g$ & 2 & $0.0473$ & $0.837$\\
$J_{\rm tot}$ & 2 & $0.2329$ & $2.851$\\
$J_q$ & 2 & $0.2135$ & $3.268$\\
$J_g$ & 2 & $0.0188$ & $0.784$\\
$B_q$ & 2 & $0.0188$ & $2.693$\\
$D_{\rm tot}$ & 3 & $-0.4211$ & $0.942$\\
$D_q$ & 3 & $-0.3739$ & $1.043$\\
$D_u$ & 3 & $-0.2067$ & $0.956$\\
$D_d$ & 3 & $-0.1757$ & $1.121$\\
$\bar c_q$ (LO + cond.) & 3 & $0.0243$ & $2.688$\\
\end{tabular}
\end{ruledtabular}
\end{table}

\begin{figure}[t]
\centering
\includegraphics[width=0.6\textwidth]{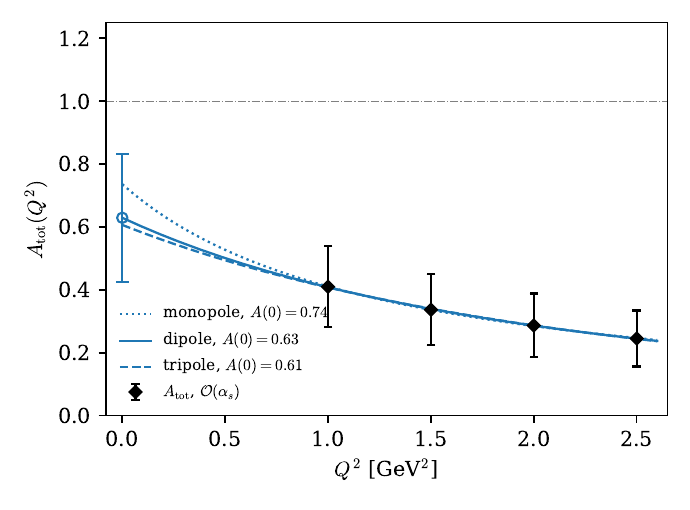}
\caption{Monopole, dipole and tripole representations of $A_{\rm tot}(Q^2)$ in $1\le Q^2\le2.5$~GeV$^2$, continued to $Q^2=0$ for illustration. The open circle is the dipole value with its uncertainty, including the spread of the three forms. The extrapolation is not constrained by our results (see text).}
\label{fig:msr}
\end{figure}

The momentum sum rule can be considered as an example. An extrapolation of $A_{\rm tot}(Q^2)$ from $1\le Q^2\le2.5$~GeV$^2$ to $Q^2=0$ (Fig.~\ref{fig:msr}) is not constrained by our results. Monopole, dipole and tripole forms describe the four values equally well, and they give $A_{\rm tot}(0)=0.74$, $0.63$ and $0.61$, respectively, all below unity. In the range $1\le Q^2\le2$~GeV$^2$, $A_{\rm tot}$ agrees with lattice QCD (Sec.~\ref{sec:compare}). However, our $Q^2$ dependence is flatter than the lattice one, and the lattice $A_{\rm tot}$ rises steeply below $Q^2=1$~GeV$^2$, a region that our analysis does not access. A possible reason for the flatter $Q^2$ dependence is the gluon-condensate contribution discussed above. A direct evaluation at small $Q^2$ is not possible in the present framework. At LO with condensates, the gluon part behaves like $1/Q^2$ over the whole range ($A_gQ^2\approx0.04$--$0.06$ for $0.1\le Q^2\le3$~GeV$^2$), and the quark part turns over below $Q^2\approx0.3$~GeV$^2$, which signals the breakdown of the local OPE at small momentum transfer.

We obtain the trace form factor $\bar c$ at LO with condensates, where $\bar c_{\rm tot}=0$ holds cell by cell and $\bar c_g=-\bar c_q$. The values of $\bar c_q$ and of $D$ at this order are given in Table~\ref{tab:cbar}. At low $Q^2$, the estimator systematic of $\bar c_q$ is of the order of one half of its value, so that only its sign and order of magnitude are determined.

\begin{table}[htb]
\caption{Protected $\bar c_q$ and $D$ at LO with condensates ($\bar c_g=-\bar c_q$, $\bar c_{\rm tot}=0$ cell by cell). The central values are averages of the two estimators, and $\sigma_{\rm est}$ is half their difference and is included in the total uncertainty.}
\label{tab:cbar}
\renewcommand{\arraystretch}{1.3}
\footnotesize
\begin{ruledtabular}
\begin{tabular}{cccccc}
$Q^2$ & $\bar c_q$ & $\sigma_{\rm est}$ & $D_q$ & $\sigma_{\rm est}$ & $D_{\rm tot}$\\
\hline
1.0 & $0.0237\pm0.0208$ & $0.0181$ & $-0.541\pm0.219$ & $0.064$ & $-0.535\pm0.211$\\
1.5 & $0.0208\pm0.0135$ & $0.0106$ & $-0.287\pm0.142$ & $0.025$ & $-0.266\pm0.138$\\
2.0 & $0.0174\pm0.0098$ & $0.0069$ & $-0.174\pm0.104$ & $0.012$ & $-0.156\pm0.102$\\
2.5 & $0.0146\pm0.0075$ & $0.0048$ & $-0.114\pm0.082$ & $0.007$ & $-0.101\pm0.081$\\
3.0 & $0.0124\pm0.0061$ & $0.0036$ & $-0.080\pm0.067$ & $0.004$ & $-0.069\pm0.066$\\
\end{tabular}
\end{ruledtabular}
\end{table}

\subsection{Comparison with other determinations}\label{sec:compare}

In the quantitative comparison, we only consider determinations that specify the renormalization scale, and we evaluate them at our values of $Q^2$ from the parametrizations given in the original works. The total form factors are scale independent and are compared directly. The quark and gluon parts are compared at $\mu=1$~GeV. When a determination is given at another scale, its quark-singlet and gluon parts are evolved with the first-order DGLAP evolution of the $n=2$ moments, which is the same evolution as in our scale variation. The lattice QCD results of Ref.~\cite{Hackett:2023rif} are given at $\mu=2$~GeV in the $\overline{\rm MS}$ scheme, at a pion mass of about $170$~MeV and for $0\le-t\le2$~GeV$^2$. We use their dipole fits and take the parameter uncertainties as independent, since their correlations are not given, and we check the dependence on the fit form with their $z$-expansion fits. For the light-front results of Ref.~\cite{Nair:2025sfr} ($\mu=2$~GeV), we use the dipole fits to the total form factors, and for the light-cone sum-rule results of Ref.~\cite{Dehghan:2025ncw} ($\mu=1$~GeV, up and down quarks), the parametrization of their results in $1\le Q^2\le10$~GeV$^2$. The comparison is presented in Tables~\ref{tab:cmp_total} and~\ref{tab:cmp_parts}.

\begin{table}[htb]
\caption{Total gravitational form factors at $\mathcal{O}(\alpha_s)$ compared with lattice QCD and with a light-front calculation, evaluated at our values of $Q^2$ (GeV$^2$) from the parametrizations of the original works. $^{\dagger}$Beyond the momentum range of the lattice data ($-t\le2$~GeV$^2$).}
\label{tab:cmp_total}
\renewcommand{\arraystretch}{1.3}
\footnotesize
\begin{ruledtabular}
\begin{tabular}{ccccc}
 & $Q^2$ & This work & Lattice QCD~\cite{Hackett:2023rif} & Light front~\cite{Nair:2025sfr}\\
\hline
$A_{\rm tot}$ & 1.0 & $0.410\pm0.129$ & $0.429\pm0.018$ & $0.273\pm0.003$\\
$A_{\rm tot}$ & 1.5 & $0.337\pm0.113$ & $0.312\pm0.015$ & $0.178\pm0.002$\\
$A_{\rm tot}$ & 2.0 & $0.287\pm0.102$ & $0.237\pm0.012$ & $0.125\pm0.002$\\
$A_{\rm tot}$ & 2.5$^{\dagger}$ & $0.245\pm0.089$ & $0.187\pm0.010$ & $0.093\pm0.001$\\
\hline
$J_{\rm tot}$ & 1.0 & $0.235\pm0.088$ & $0.243\pm0.018$ & $0.135\pm0.002$\\
$J_{\rm tot}$ & 1.5 & $0.206\pm0.100$ & $0.183\pm0.016$ & $0.088\pm0.001$\\
$J_{\rm tot}$ & 2.0 & $0.188\pm0.116$ & $0.143\pm0.014$ & $0.062\pm0.001$\\
$J_{\rm tot}$ & 2.5$^{\dagger}$ & $0.176\pm0.133$ & $0.115\pm0.013$ & $0.046\pm0.001$\\
\hline
$D_{\rm tot}$ & 1.0 & $-0.423\pm0.201$ & $-0.334\pm0.136$ & $-0.373\pm0.075$\\
$D_{\rm tot}$ & 1.5 & $-0.206\pm0.121$ & $-0.188\pm0.086$ & $-0.207\pm0.041$\\
$D_{\rm tot}$ & 2.0 & $-0.117\pm0.083$ & $-0.120\pm0.059$ & $-0.131\pm0.026$\\
$D_{\rm tot}$ & 2.5$^{\dagger}$ & $-0.074\pm0.062$ & $-0.084\pm0.045$ & $-0.091\pm0.018$\\
\end{tabular}
\end{ruledtabular}
\end{table}

\begin{table}[htb]
\caption{Quark ($u+d$) and gluon parts at $\mu=1$~GeV in the $\overline{\rm MS}$ scheme. The lattice values~\cite{Hackett:2023rif} are evolved from $\mu=2$~GeV with the first-order DGLAP evolution of the $n=2$ moments. $^{\dagger}$As in Table~\ref{tab:cmp_total}.}
\label{tab:cmp_parts}
\renewcommand{\arraystretch}{1.3}
\footnotesize\setlength{\tabcolsep}{3pt}
\begin{ruledtabular}
\begin{tabular}{ccccccc}
 & & \multicolumn{3}{c}{quark ($u+d$)} & \multicolumn{2}{c}{gluon}\\
 & $Q^2$ & This work & Lattice~\cite{Hackett:2023rif} & LCSR~\cite{Dehghan:2025ncw} & This work & Lattice~\cite{Hackett:2023rif}\\
\hline
A & 1.0 & $0.364\pm0.120$ & $0.256\pm0.012$ & $0.239\pm0.008$ & $0.047\pm0.029$ & $0.167\pm0.012$\\
A & 1.5 & $0.309\pm0.110$ & $0.192\pm0.010$ & $0.151\pm0.006$ & $0.029\pm0.024$ & $0.116\pm0.009$\\
A & 2.0 & $0.269\pm0.101$ & $0.150\pm0.008$ & $0.101\pm0.004$ & $0.019\pm0.021$ & $0.085\pm0.007$\\
A & 2.5$^{\dagger}$ & $0.233\pm0.090$ & $0.120\pm0.007$ & $0.070\pm0.003$ & $0.013\pm0.018$ & $0.065\pm0.005$\\
\hline
J & 1.0 & $0.215\pm0.086$ & $0.141\pm0.013$ & $0.085\pm0.003$ & $0.019\pm0.016$ & $0.100\pm0.009$\\
J & 1.5 & $0.194\pm0.100$ & $0.109\pm0.012$ & $0.053\pm0.002$ & $0.011\pm0.014$ & $0.073\pm0.007$\\
J & 2.0 & $0.180\pm0.116$ & $0.087\pm0.011$ & $0.035\pm0.001$ & $0.007\pm0.013$ & $0.055\pm0.006$\\
J & 2.5$^{\dagger}$ & $0.171\pm0.133$ & $0.071\pm0.010$ & $0.024\pm0.001$ & $0.005\pm0.013$ & $0.043\pm0.005$\\
\hline
D & 1.0 & $-0.375\pm0.189$ & $-0.215\pm0.093$ & $-0.376\pm0.036$ & $-0.041\pm0.065$ & $-0.108\pm0.075$\\
D & 1.5 & $-0.195\pm0.118$ & $-0.129\pm0.061$ & $-0.201\pm0.024$ & $-0.007\pm0.028$ & $-0.054\pm0.042$\\
D & 2.0 & $-0.115\pm0.082$ & $-0.086\pm0.043$ & $-0.120\pm0.016$ & $0.000\pm0.016$ & $-0.032\pm0.027$\\
D & 2.5$^{\dagger}$ & $-0.075\pm0.061$ & $-0.062\pm0.033$ & $-0.077\pm0.012$ & $0.002\pm0.010$ & $-0.021\pm0.019$\\
\end{tabular}
\end{ruledtabular}
\end{table}

The total form factors agree with the lattice results within $0.7\sigma$ at all values of $Q^2$, for $A$, $J$ and $D$ alike, and $D_{\rm tot}$ also agrees with the light-front result. The light-front values of $A_{\rm tot}$ and $J_{\rm tot}$ are smaller and fall faster with $Q^2$ than both our results and the lattice results. Our $A_{\rm tot}$ and $J_{\rm tot}$, on the other hand, fall more slowly than the lattice ones. This may be related to the gluon-condensate contribution of Table~\ref{tab:opecomp} and, for $J_{\rm tot}$, also to the four-quark contributions at $\beta\neq1$ (Sec.~\ref{sec:numerics}). The quark part of $D$ agrees with both the lattice and the light-cone sum-rule results, and the gluon part of $D$ is compatible with the lattice result within about $1\sigma$. A larger difference, of $2.7$--$4.5\sigma$, appears in the gluon shares of $A$ and $J$, which are smaller than the lattice values by factors of $3.6$--$5$ for $A_g$ and $5$--$9$ for $J_g$. The quark parts are correspondingly larger, by $0.7$--$1.3\sigma$, so that the totals agree. A possible reason is that at this order the gluon EMT couples to the quark lines only through one-gluon exchange and through the gluon condensate, while the gluon momentum fraction of the nucleon at low scales is to a large extent of non-perturbative origin. We have not verified this explanation. The light-cone sum-rule values of $A_q$ and $J_q$ are smaller than ours and than the evolved lattice values, and they fall faster with $Q^2$. The values of $\bar c_q$ at LO with condensates (Table~\ref{tab:cbar}) are positive and of the same size as those of Ref.~\cite{Dehghan:2025ncw} at $\mu=1$~GeV, which are $0.040$, $0.029$, $0.021$ and $0.016$ at $Q^2=1$, $1.5$, $2$ and $2.5$~GeV$^2$, respectively.

The experimental extractions of the quark $D$-term~\cite{Burkert:2018bqq} and of the gluon GFFs~\cite{Duran:2022xag} refer to momentum transfers that only partly overlap with our range, and they rely on model frameworks for the $t$ dependence. In addition, the renormalization scale is not specified, and therefore we do not include them in the quantitative comparison. The same applies to model calculations without a definite renormalization scale, which have been mentioned in Sec.~\ref{sec:intro}.

\section{Limitations}\label{sec:limitations}

In the present extraction, the transitions between the nucleon and excited states are treated only in part. With equal Borel masses and a Borel window of finite width, they add a term $C\,e^{-\Delta/M^2}$ to the normalized correlator, $H(M^2)=F+C\,e^{-\Delta/M^2}$, and $F$ and $C$ cannot be separated. A fit of this form attributes all $M^2$ dependence to transitions and amplifies it, and in an exactly conserved leading-order sector it shifts the extracted values by amounts comparable to the values themselves. The part of the transition contribution that varies with $M^2$ is contained in the spread over the working region, since the central value and the spread are taken over all cells along $M^2$. In fixed rows of $(\beta,s_0,S_0)$, the half-range along $M^2$ is $0.04$--$0.39$ (median) and at most $0.73$ of the full spread, over all form factors and flavours at $\mathcal{O}(\alpha_s)$ and at LO with condensates. The $\mathcal{O}(\alpha_s)$ values for the main ones are given in Table~\ref{tab:transitions}. A constant offset over the window, $C\langle e^{-\Delta/M^2}\rangle$, cannot be determined and is not included. Unequal Borel masses do not change this within the present criteria. Each Borel variable must lie in the two-point window, $0.67\le M^2\le0.95$~GeV$^2$, where $e^{-\Delta/M^2}$ varies only between $0.17$ and $0.29$ for $\Delta=m_{N(1440)}^2-m_N^2$. Therefore, the transition term cannot be separated from a constant with one or with two Borel variables, and an $M^2$ dependence of other origin would enter the extracted constant amplified by factors of $1.6$--$3.4$, depending on $\Delta$ and on the width of the window. Its determination requires a wider window or external information on the transitions, such as the gravitational transition form factors between the nucleon and its resonances~\cite{Ozdem:2019pkg,Ozdem:2022zig}.

\begin{table}[htb]
\caption{Fraction of the working-region spread due to the variation along $M^2$, $\sigma_{M^2}/\sigma_{\rm region}$ (median / maximum over fixed $(\beta,s_0,S_0)$ rows), at $\mathcal{O}(\alpha_s)$.}
\label{tab:transitions}
\renewcommand{\arraystretch}{1.3}
\small
\begin{ruledtabular}
\begin{tabular}{lcccc}
 & $Q^2=1$ & $1.5$ & $2$ & $2.5$\\
\hline
$A_{\rm tot}$ & 0.12 / 0.35 & 0.12 / 0.36 & 0.13 / 0.37 & 0.14 / 0.26\\
$J_{\rm tot}$ & 0.08 / 0.19 & 0.06 / 0.20 & 0.06 / 0.22 & 0.04 / 0.23\\
$B_q$ & 0.35 / 0.57 & 0.34 / 0.56 & 0.32 / 0.52 & 0.36 / 0.38\\
$D_{\rm tot}$ & 0.14 / 0.31 & 0.11 / 0.28 & 0.09 / 0.23 & 0.08 / 0.21\\
$D_g$ & 0.17 / 0.55 & 0.17 / 0.53 & 0.15 / 0.51 & 0.15 / 0.45\\
\end{tabular}
\end{ruledtabular}
\end{table}

The gluon parts of $A$ and $J$ are smaller than in lattice QCD (Sec.~\ref{sec:compare}). At this order, the gluon EMT couples to the quark lines only through one-gluon exchange and through the gluon condensate, and higher orders may be required to describe the gluon momentum fraction.

The $Q^2$ dependence of $A$ and $J$ is affected by the gluon-condensate contribution to the quark-EMT correlator, which is the largest one in their correlators and depends only weakly on $Q^2$ (Table~\ref{tab:opecomp}). This is a feature of the local OPE, and a treatment beyond the local approximation, for instance with a nonlocal gluon condensate~\cite{Mikhailov:1991pt,Bakulev:1991ps}, may modify the $Q^2$ dependence of these form factors.

The local OPE also restricts the analysis to $Q^2\gtrsim1$~GeV$^2$, as shown by the small-$Q^2$ behaviour described above. The forward values cannot be obtained from our results without an extrapolation that the results do not constrain.

For the same reason, we do not extract the pressure and shear distributions or the mechanical radius, which require $D(t)$ at small momentum transfer and its forward value.

The $\mathcal{O}(\alpha_s)$ determination of $\bar c$ requires the $S_7$ structure. Its $\mathcal{O}(\alpha_s)$ coefficient has been computed at one kinematic point, where the individual conservation identities deviate from zero at the level of two standard deviations (Sec.~\ref{sec:validation}). Therefore, we give $\bar c$ at LO with condensates. Whether a finite term is missing can be settled either by a higher numerical precision of the antisymmetric moments, at more than one kinematic point, or by an identification of the term itself.

The quark and gluon parts are $\overline{\rm MS}$ quantities at $\mu=1$~GeV. The cancellation of the structure-preserving prescription for the current subgraphs is supported by the forward Ward identity, but we have not tested it with a second prescription.

The results also refer to a particular class of interpolating currents. The working region exists only for $\beta$ close to $1$ (Sec.~\ref{sec:numerics}), so that the dependence of the results on the current, and in particular a comparison with the Ioffe current, cannot be tested within the present analysis.

For the smaller quantities ($A_d$, $A_g$, $B_q$, $B_d$, $B_g$, $B_{\rm tot}$, $D_g$) and for $D_{\rm tot}$, the numerical uncertainty of the $\mathcal{O}(\alpha_s)$ integrals is not negligible relative to the other uncertainties. It amounts to up to $\sim50\%$ for $D_g$ and $12$--$18\%$ for $D_{\rm tot}$. This uncertainty is dominated by the statistical error of the antisymmetric moments, and it is included in the quoted uncertainties.

\section{Conclusions}\label{sec:conclusions}

The gravitational form factors of the nucleon carry information on the decomposition of its mass and spin into quark and gluon contributions and on its internal forces. In this study, we have computed the form factors $A$, $J$, $B$ and $D$ of the nucleon with the help of three-point QCD sum rules, including the perturbative contribution at $\mathcal{O}(\alpha_s)$ for both the quark and the gluon energy-momentum tensor, in the $\overline{\rm MS}$ scheme at $\mu=1$~GeV. The form factors are extracted from combinations of Lorentz structures that are protected by the Ward identities. For $D$ and $\bar c$, we have shown that estimators built from equal Borel masses alone cannot be protected by conservation. Instead, we have constructed protected estimators from the conservation identities in Borel space, which involve antisymmetric Borel moments. Since the $D$ estimators do not depend on the $g_{\mu\nu}$ structure, $D$ can be obtained at $\mathcal{O}(\alpha_s)$, although $\bar c$ remains at leading order with condensates.

The $\mathcal{O}(\alpha_s)$ three-point functions satisfy the forward Ward identity for $A$ and $J$ within $1.5\%$, the renormalization-group constraint on their scale dependence, and the conservation identity $V_1+V_3=0$ at seven kinematic points. Compared with the leading-order analysis with condensates, the $\mathcal{O}(\alpha_s)$ corrections reduce $A_{\rm tot}$ by factors of $0.60$--$0.68$, $J_{\rm tot}$ by factors of $0.73$--$0.78$, and $|D_{\rm tot}|$ by $21$--$27\%$. In the range $1\le Q^2\le2$~GeV$^2$, the total form factors agree with lattice QCD within $0.7\sigma$, and the quark and gluon parts of $D$ within about $1\sigma$. The gluon parts of $A$ and $J$ are smaller than the lattice values, and the quark parts are correspondingly larger. In the working region, the perturbative contribution is the largest one for $D$ up to $Q^2=2$~GeV$^2$. For $A$ and $J$, however, the largest one is the gluon-condensate contribution to the quark-EMT correlator, which depends only weakly on $Q^2$, and the forward values are not constrained by our results.

The main limitations of the present analysis are the treatment of transitions to excited states, whose $M^2$-independent part cannot be determined within the Borel windows of the sum rules, the restriction to $Q^2\ge1$~GeV$^2$ by the local OPE, the large and weakly $Q^2$-dependent gluon-condensate contribution to $A$ and $J$, and the trace form factor $\bar c$, whose $\mathcal{O}(\alpha_s)$ determination requires the $g_{\mu\nu}$ structure. External information on the transitions to excited states, a treatment of the gluon condensate beyond the local approximation and the resolution of the $g_{\mu\nu}$ structure at $\mathcal{O}(\alpha_s)$ may address these limitations. The mechanical properties, which require $D(t)$ at small momentum transfer, are left for such an extension. Our results can be compared with future lattice QCD calculations of the gluon form factors and with experimental extractions at larger momentum transfer.

\section*{USE of AI}
The AI tools (Claude Opus 5.5) were used to assist with writing the numerical code, drawing the figures, and language editing and improving the readability of the manuscript. The research questions, the methodology, and the physics decisions are the author's. All analytical analyses were obtained by the author, and the results were independently verified with the author's own Mathematica code. The author directed the AI-assisted work and takes full responsibility for the content of the manuscript.

\appendix
\section{OPE expressions}\label{app:ope}

\subsection{Representation and double Borel transformation}

At LO we write the coefficient of each structure $S_i$ in Feynman-parameter form,
\begin{equation}
\begin{gathered}
\Pi_i(p,p')=\sum_{k,l}\int_{\Delta}du\,dw\;c_{i;kl}(u,w)\,X^k\,(\ln X)^l,\\
X=-\big(uv\,q^2+vw\,p^2+uw\,p'^2\big),\qquad v=1-u-w,
\end{gathered}
\label{eq:feynrep}
\end{equation}
where $\Delta$ is the simplex $u,w\ge0$, $u+w\le1$. The coefficient functions depend on $\beta$, on the condensates and on $p\cdot p'$, $p^2$ and $p'^2$. There are logarithmic terms ($l=1$, $k\ge0$) and pole terms ($l=0$, $k\le-1$), and the condensate contributions also contain factorized terms of the form $c\,X_x^{k_x}(\ln X_x)^{l_x}X_y^{k_y}(\ln X_y)^{l_y}$ with $X_x=-p^2$ and $X_y=-p'^2$.

We normalize the double Borel transformation by $\hat B[1/(s-p^2)]=e^{-s\tau_1}$ and $\hat B[1/(s'-p'^2)]=e^{-s'\tau_2}$. After the replacement $p\cdot p'=(p^2+p'^2-q^2)/2$, factors of $p^2$ and $p'^2$ in the coefficients turn into $-\partial_{\tau_1}$ and $-\partial_{\tau_2}$. For a coefficient that does not depend on $p^2$ and $p'^2$, the Schwinger representation of $X^k$ gives, with $T=\tau_1+\tau_2$ and $Q^2=-q^2$,
\begin{align}
\hat B\big[c\,X^k\ln X\big]&=(-1)^{k+1}k!\;T^{-k-2}\int_0^1dw\;w^k(1-w)^{k+1}\,c(u_B,w)\,E(w),\label{eq:borellog}\\
\hat B\big[c\,X^{-a}\big]&=\frac{1}{\Gamma(a)}\int_0^1dw\;\frac{\lambda^{a-2}}{w^2(1-w)}\,c(u_B,w)\,E(w),\label{eq:borelpole}
\end{align}
where $u_B=\tau_2(1-w)/T$, $\lambda=T/[w(1-w)]$ and $E(w)=\exp[-Q^2\tau_1\tau_2(1-w)/(Tw)]$. Terms that depend on only one of $p^2$ and $p'^2$ vanish. For the factorized terms, $X^n\ln X\to-(-1)^n n!/\tau^{n+1}$ ($n\ge0$), $X^{-k}\to\tau^{k-1}/(k-1)!$, and polynomials vanish, separately in each variable.

In the analysis we subtract the continuum in the double spectral representation,
\begin{equation}
\hat B\to\int\!\!\int_{s+s'\le S_0}ds\,ds'\,\rho(s,s')\,e^{-\tau_1s-\tau_2s'},
\end{equation}
where $\rho$ follows from the discontinuity of the integrand of Eq.~\eqref{eq:feynrep} across the root of $X=0$. We have checked Eqs.~\eqref{eq:borellog} and~\eqref{eq:borelpole} and the rules for the factorized terms independently against the numerical double-dispersion integral at $S_0=\infty$, for all types of terms in the perturbative and condensate contributions of several structures. The agreement is $10^{-6}$--$10^{-8}$ where the coefficients contain $p^2$ or $p'^2$ (finite-difference derivatives), $10^{-11}$ or better otherwise, and exact for the factorized terms.

\subsection{Two-point function}

The residue follows from the coefficient of $\slashed q$ in the two-point function of the current~\eqref{eq:current},
\begin{equation}
\begin{split}
\lambda^2e^{-m_N^2/M^2}={}&\frac{5+2\beta+5\beta^2}{512\pi^4}\int_0^{s_0}ds\,s^2e^{-s/M^2}\Big[1+a\Big(c_0+3\ln\frac{\mu^2}{s}\Big)\Big]
+\frac{5+2\beta+5\beta^2}{1024\pi^4}\,\langle g^2G^2\rangle\int_0^{s_0}ds\,e^{-s/M^2}\\
&+\frac{(\beta-1)(7\beta+5)}{6}\,\langle\bar qq\rangle^2-\frac{(\beta-1)(13\beta+11)}{24M^2}\,m_0^2\langle\bar qq\rangle^2,
\end{split}
\label{eq:twopoint}
\end{equation}
with $a=\alpha_sC_F/(4\pi)$ and $c_0=22.75$, independent of $\beta$, from our $\mathcal{O}(\alpha_s)$ calculation.

\subsection{Perturbative contribution}
For each structure $S_i$ and quark flavour the LO perturbative coefficient is
\begin{equation}
\Pi^{(0)}_{i,f}(p,p')=\int_{\Delta}du\,dw\;\sum_{k}c^{(0)}_{i,f;k}(u,w)\,X^k\ln X ,
\end{equation}
with the coefficient functions listed below, organized by powers of $\beta$ and by monomials in $p\cdot p'$, $p^2$ and $p'^2$. The brackets depend on $u$, $w$ and $v=1-u-w$ only.
\begin{align*}
c^{(0)}_{S_1,u}\big|_{\ln X}&=\left(\beta + 1\right)^{2}\Big\{ (p\cdot p^{\prime})\,\Big[\frac{u v^{2} \left(v - 1\right) \left(2 v - 1\right)}{16 \pi^{4} w}\Big] +p^{2}\,\Big[\frac{v^{3} \left(v - 1\right) \left(2 v - 1\right)}{32 \pi^{4} w}\Big]\\
&\quad+p^{\prime 2}\,\Big[\frac{u^{2} v \left(v - 1\right) \left(2 v - 1\right)}{32 \pi^{4} w}\Big]\Big\}\\
c^{(0)}_{S_1,u}\big|_{X\ln X}&=\left(\beta^{2} + 1\right) \Big[\frac{v}{32 \pi^{4} w^{2}}\left(10 u^{2} w - 10 u^{2} + 20 u w^{2} - 28 u w + 2 u + 10 w^{3} - 18 w^{2} + 3 w + 1\right)\Big] +\beta\\
&\quad\Big[\frac{v}{16 \pi^{4} w^{2}}\left(10 u^{2} w - 4 u^{2} + 20 u w^{2} - 16 u w + 2 u + 10 w^{3} - 12 w^{2} + 3 w\right)\Big]\\
c^{(0)}_{S_1,d}\big|_{\ln X}&=\left(\beta + 1\right)^{2}\Big\{ p^{\prime 2}\,\Big[\frac{u v \left(2 v - 1\right)}{64 \pi^{4} w}\Big]\Big\}\\
c^{(0)}_{S_1,d}\big|_{X\ln X}&=\left(\beta^{2} + 1\right) \Big[\frac{v}{64 \pi^{4} w^{2}}\left(14 u^{2} + 28 u w - 13 u + 14 w^{2} - 13 w + 2\right)\Big] +\beta \Big[\frac{v \left(v - 1\right) \left(2 v - 1\right)}{32 \pi^{4} w^{2}}\Big]\\
c^{(0)}_{S_2,u}\big|_{\ln X}&=\left(\beta + 1\right)^{2}\Big\{ (p\cdot p^{\prime})\,\Big[- \frac{u v \left(2 v - 1\right) \left(2 u^{2} + 2 u w - 2 u - w\right)}{32 \pi^{4} w}\Big]\\
&\quad+p^{2}\,\Big[- \frac{v^{2} \left(2 v - 1\right) \left(2 u^{2} + 2 u w - 2 u - w\right)}{64 \pi^{4} w}\Big]\\
&\quad+p^{\prime 2}\,\Big[- \frac{u^{2} \left(2 v - 1\right) \left(2 u^{2} + 2 u w - 2 u - w\right)}{64 \pi^{4} w}\Big]\Big\}\\
c^{(0)}_{S_2,u}\big|_{X\ln X}&=\left(\beta^{2} + 1\right) \Big[\frac{1}{64 \pi^{4} w^{2}}\Big(\\
&\quad20 u^{3} w - 20 u^{3} + 40 u^{2} w^{2} - 72 u^{2} w + 36 u^{2} + 20 u w^{3} - 60 u w^{2} + 66 u w - 14 u - 8 w^{3}\\
&\quad+23 w^{2} - 9 w\Big)\Big] +\beta \Big[\frac{1}{32 \pi^{4} w^{2}}\Big(\\
&\quad20 u^{3} w - 8 u^{3} + 40 u^{2} w^{2} - 48 u^{2} w + 12 u^{2} + 20 u w^{3} - 48 u w^{2} + 30 u w - 4 u - 8 w^{3}\\
&\quad+11 w^{2} - 3 w\Big)\Big]\\
c^{(0)}_{S_2,d}\big|_{\ln X}&=\left(\beta + 1\right)^{2}\Big\{ (p\cdot p^{\prime})\,\Big[- \frac{u v \left(2 u + w\right) \left(2 v - 1\right)}{64 \pi^{4} w}\Big] +p^{2}\,\Big[- \frac{v^{2} \left(2 u + w\right) \left(2 v - 1\right)}{128 \pi^{4} w}\Big]\\
&\quad+p^{\prime 2}\,\Big[- \frac{u^{2} \left(2 v - 1\right) \left(2 u + w - 2\right)}{128 \pi^{4} w}\Big]\Big\}\\
c^{(0)}_{S_2,d}\big|_{X\ln X}&=\left(\beta^{2} + 1\right) \Big[\frac{1}{128 \pi^{4} w^{2}}\left(28 u^{3} + 72 u^{2} w - 30 u^{2} + 52 u w^{2} - 46 u w + 6 u + 8 w^{3} - 11 w^{2} + 3 w\right)\Big] +\beta\\
&\quad\Big[\frac{1}{64 \pi^{4} w^{2}}\left(4 u^{3} + 24 u^{2} w - 6 u^{2} + 28 u w^{2} - 22 u w + 2 u + 8 w^{3} - 11 w^{2} + 3 w\right)\Big]\\
c^{(0)}_{S_5,u}\big|_{\ln X}&=\left(\beta + 1\right)^{2}\Big\{ (p\cdot p^{\prime})\,\Big[\frac{u v}{64 \pi^{4} w}\left(8 u^{3} + 16 u^{2} w - 12 u^{2} + 8 u w^{2} - 16 u w + 4 u - 4 w^{2} + 3 w\right)\Big]\\
&\quad+p^{2}\,\Big[\frac{v^{2}}{128 \pi^{4} w}\left(8 u^{3} + 16 u^{2} w - 12 u^{2} + 8 u w^{2} - 16 u w + 4 u - 4 w^{2} + 3 w\right)\Big]\\
&\quad+p^{\prime 2}\,\Big[\frac{u^{2}}{128 \pi^{4} w}\left(8 u^{3} + 16 u^{2} w - 12 u^{2} + 8 u w^{2} - 16 u w + 4 u - 4 w^{2} + 3 w\right)\Big]\Big\}\\
c^{(0)}_{S_5,u}\big|_{X\ln X}&=\left(\beta^{2} + 1\right) \Big[\frac{1}{128 \pi^{4} w^{2}}\Big(\\
&\quad40 u^{3} w - 40 u^{3} + 80 u^{2} w^{2} - 144 u^{2} w + 48 u^{2} + 40 u w^{3} - 120 u w^{2} + 96 u w - 4 u\\
&\quad- 16 w^{3} + 37 w^{2} - 11 w - 4\Big)\Big] +\beta \Big[\frac{1}{64 \pi^{4} w^{2}}\Big(\\
&\quad40 u^{3} w - 16 u^{3} + 80 u^{2} w^{2} - 96 u^{2} w + 24 u^{2} + 40 u w^{3} - 96 u w^{2} + 60 u w - 8 u - 16 w^{3}\\
&\quad+25 w^{2} - 9 w\Big)\Big]\\
c^{(0)}_{S_5,d}\big|_{\ln X}&=\left(\beta + 1\right)^{2}\Big\{ (p\cdot p^{\prime})\,\Big[- \frac{u v \left(1 - 2 v\right) \left(2 u + w - 2\right)}{128 \pi^{4} w}\Big] +p^{2}\,\Big[- \frac{v^{2} \left(1 - 2 v\right) \left(2 u + w\right)}{256 \pi^{4} w}\Big]\\
&\quad+p^{\prime 2}\,\Big[- \frac{u}{256 \pi^{4} w}\left(4 u^{3} + 6 u^{2} w - 2 u^{2} + 2 u w^{2} - u w - 4 u - 2 w + 2\right)\Big]\Big\}\\
c^{(0)}_{S_5,d}\big|_{X\ln X}&=\left(\beta^{2} + 1\right)\\
&\quad\Big[\frac{1}{256 \pi^{4} w^{2}}\left(56 u^{3} + 96 u^{2} w - 108 u^{2} + 32 u w^{2} - 118 u w + 60 u - 8 w^{3} - 15 w^{2} + 31 w - 8\right)\Big] +\beta\\
&\quad\Big[\frac{1}{128 \pi^{4} w^{2}}\left(8 u^{3} - 12 u^{2} - 16 u w^{2} + 2 u w + 4 u - 8 w^{3} + 9 w^{2} - w\right)\Big]\\
c^{(0)}_{S_7,u}\big|_{X\ln X}&=\left(\beta + 1\right)^{2}\Big\{ (p\cdot p^{\prime})\,\Big[\frac{u v \left(2 v - 1\right)}{32 \pi^{4} w}\Big] +p^{2}\,\Big[\frac{v^{2} \left(2 v - 1\right)}{64 \pi^{4} w}\Big] +p^{\prime 2}\,\Big[\frac{u^{2} \left(2 v - 1\right)}{64 \pi^{4} w}\Big]\Big\}\\
c^{(0)}_{S_7,u}\big|_{X^2\ln X}&=\left(\beta^{2} + 1\right) \Big[\frac{1}{128 \pi^{4} u w^{2}}\left(- 8 u^{2} w + 10 u^{2} - 8 u w^{2} + 15 u w - 2 u + 4 w - 1\right)\Big] +\beta\\
&\quad\Big[- \frac{8 u w - 4 u + 8 w^{2} - 9 w + 2}{64 \pi^{4} w^{2}}\Big]\\
c^{(0)}_{S_7,d}\big|_{X\ln X}&=\left(\beta + 1\right)^{2}\Big\{ (p\cdot p^{\prime})\,\Big[\frac{v \left(u - 1\right)}{64 \pi^{4} w}\Big] +p^{2}\,\Big[\frac{v^{2}}{128 \pi^{4} w}\Big] +p^{\prime 2}\,\Big[\frac{u^{2}}{128 \pi^{4} w}\Big]\Big\}\\
c^{(0)}_{S_7,d}\big|_{X^2\ln X}&=\left(\beta^{2} + 1\right) \Big[- \frac{14 u^{2} + 11 u w - 13 u - 8 w + 2}{256 \pi^{4} u w^{2}}\Big] +\beta \Big[- \frac{2 u - w - 1}{128 \pi^{4} w^{2}}\Big]\\
c^{(0)}_{S_9,u}\big|_{\ln X}&=\left(\beta + 1\right)^{2}\Big\{ (p\cdot p^{\prime})^{2}\,\Big[\frac{u v \left(2 v - 1\right) \left(2 u^{2} + 2 u w - 2 u - w\right)}{64 \pi^{4} w}\Big]\\
&\quad+(p\cdot p^{\prime})\,p^{2}\,\Big[\frac{v^{2} \left(2 v - 1\right) \left(4 u^{2} + 4 u w - 2 u - w\right)}{128 \pi^{4} w}\Big]\\
&\quad+(p\cdot p^{\prime})\,p^{\prime 2}\,\Big[\frac{u^{2} \left(2 v - 1\right) \left(4 u^{2} + 4 u w - 6 u - 3 w + 2\right)}{128 \pi^{4} w}\Big] +(p^{2})^{2}\,\Big[\frac{v^{3} \left(1 - v\right) \left(2 v - 1\right)}{128 \pi^{4} w}\Big]\\
&\quad+p^{2}\,p^{\prime 2}\,\Big[\frac{u v \left(2 v - 1\right) \left(2 u^{2} + 2 u w - 2 u - w + 1\right)}{128 \pi^{4} w}\Big] +(p^{\prime 2})^{2}\,\Big[- \frac{u^{3} \left(u - 1\right) \left(2 v - 1\right)}{128 \pi^{4} w}\Big]\Big\}\\
c^{(0)}_{S_9,u}\big|_{X\ln X}&=\left(\beta^{2} + 1\right)\Big\{ (p\cdot p^{\prime})\,\Big[\frac{1}{128 \pi^{4} w^{2}}\Big(\\
&\quad- 24 u^{3} w + 32 u^{3} - 48 u^{2} w^{2} + 102 u^{2} w - 48 u^{2} - 24 u w^{3} + 78 u w^{2} - 80 u w + 16 u\\
&\quad+8 w^{3} - 23 w^{2} + 9 w\Big)\Big]\\
&\quad+p^{2}\,\Big[- \frac{v}{128 \pi^{4} w^{2}}\left(12 u^{2} w - 16 u^{2} + 24 u w^{2} - 43 u w + 8 u + 12 w^{3} - 27 w^{2} + 10 w\right)\Big]\\
&\quad+p^{\prime 2}\,\Big[\frac{u}{128 \pi^{4} w^{2}}\left(12 u^{2} w - 16 u^{2} + 12 u w^{2} - 31 u w + 24 u - 8 w^{2} + 21 w - 8\right)\Big]\Big\} +\beta\Big\{\\
&\quad(p\cdot p^{\prime})\,\Big[\frac{1}{64 \pi^{4} w^{2}}\Big(\\
&\quad- 24 u^{3} w + 8 u^{3} - 48 u^{2} w^{2} + 54 u^{2} w - 12 u^{2} - 24 u w^{3} + 54 u w^{2} - 32 u w + 4 u + 8 w^{3}\\
&\quad- 11 w^{2} + 3 w\Big)\Big]\\
&\quad+p^{2}\,\Big[- \frac{v}{64 \pi^{4} w^{2}}\left(12 u^{2} w - 4 u^{2} + 24 u w^{2} - 19 u w + 2 u + 12 w^{3} - 15 w^{2} + 4 w\right)\Big]\\
&\quad+p^{\prime 2}\,\Big[\frac{u}{64 \pi^{4} w^{2}}\left(12 u^{2} w - 4 u^{2} + 12 u w^{2} - 19 u w + 6 u - 8 w^{2} + 9 w - 2\right)\Big]\Big\}\\
c^{(0)}_{S_9,u}\big|_{X^2\ln X}&=\left(\beta^{2} + 1\right) \Big[\frac{8 u w - 28 u + 8 w^{2} - 33 w + 14}{256 \pi^{4} w^{2}}\Big] +\beta \Big[\frac{8 u w - 4 u + 8 w^{2} - 9 w + 2}{128 \pi^{4} w^{2}}\Big]\\
c^{(0)}_{S_9,d}\big|_{\ln X}&=\left(\beta + 1\right)^{2}\Big\{ (p\cdot p^{\prime})^{2}\,\Big[- \frac{u v \left(2 v - 1\right) \left(2 u^{2} + 2 u w - 2 u - w\right)}{128 \pi^{4} w}\Big]\\
&\quad+(p\cdot p^{\prime})\,p^{2}\,\Big[- \frac{v^{2} \left(2 v - 1\right) \left(4 u^{2} + 4 u w - 2 u - w\right)}{256 \pi^{4} w}\Big]\\
&\quad+(p\cdot p^{\prime})\,p^{\prime 2}\,\Big[- \frac{u^{2} \left(2 v - 1\right) \left(4 u^{2} + 4 u w - 6 u - 3 w + 2\right)}{256 \pi^{4} w}\Big]\\
&\quad+(p^{2})^{2}\,\Big[- \frac{v^{3} \left(1 - v\right) \left(2 v - 1\right)}{256 \pi^{4} w}\Big] +p^{2}\,p^{\prime 2}\,\Big[- \frac{u v \left(2 v - 1\right) \left(2 u^{2} + 2 u w - 2 u - w + 1\right)}{256 \pi^{4} w}\Big]\\
&\quad+(p^{\prime 2})^{2}\,\Big[\frac{u^{3} \left(u - 1\right) \left(2 v - 1\right)}{256 \pi^{4} w}\Big]\Big\}\\
c^{(0)}_{S_9,d}\big|_{X\ln X}&=\left(\beta + 1\right)^{2}\Big\{ (p\cdot p^{\prime})\,\Big[\frac{1}{256 \pi^{4} w^{2}}\Big(\\
&\quad32 u^{3} w - 4 u^{3} + 64 u^{2} w^{2} - 56 u^{2} w + 6 u^{2} + 32 u w^{3} - 60 u w^{2} + 28 u w - 2 u - 8 w^{3}\\
&\quad+11 w^{2} - 3 w\Big)\Big] +p^{2}\,\Big[\frac{v}{256 \pi^{4} w^{2}}\left(16 u^{2} w - 2 u^{2} + 32 u w^{2} - 20 u w + u + 16 w^{3} - 18 w^{2} + 4 w\right)\Big]\\
&\quad+p^{\prime 2}\,\Big[- \frac{u \left(2 u - 1\right) \left(8 u w - u + 8 w^{2} - 5 w + 1\right)}{256 \pi^{4} w^{2}}\Big]\Big\}\\
c^{(0)}_{S_9,d}\big|_{X^2\ln X}&=\left(\beta^{2} + 1\right) \Big[\frac{\left(2 v - 1\right) \left(12 w + 11\right)}{512 \pi^{4} w^{2}}\Big] +\beta \Big[\frac{\left(2 v - 1\right) \left(12 w - 1\right)}{256 \pi^{4} w^{2}}\Big]\\
c^{(0)}_{S_{11},u}\big|_{\ln X}&=\left(\beta + 1\right)^{2}\Big\{ (p\cdot p^{\prime})\,\Big[\frac{u v \left(2 v - 1\right)}{64 \pi^{4}}\Big] +p^{2}\,\Big[\frac{v^{2} \left(2 v - 1\right)}{128 \pi^{4}}\Big] +p^{\prime 2}\,\Big[\frac{u^{2} \left(2 v - 1\right)}{128 \pi^{4}}\Big]\Big\}\\
c^{(0)}_{S_{11},u}\big|_{X\ln X}&=\left(\beta^{2} + 1\right) \Big[- \frac{8 u w - 18 u + 8 w^{2} - 23 w + 9}{128 \pi^{4} w}\Big] +\beta \Big[- \frac{8 u w - 6 u + 8 w^{2} - 11 w + 3}{64 \pi^{4} w}\Big]\\
c^{(0)}_{S_{11},d}\big|_{\ln X}&=\left(\beta + 1\right)^{2}\Big\{ (p\cdot p^{\prime})\,\Big[- \frac{u v \left(2 v - 1\right)}{128 \pi^{4}}\Big] +p^{2}\,\Big[- \frac{v^{2} \left(2 v - 1\right)}{256 \pi^{4}}\Big] +p^{\prime 2}\,\Big[- \frac{u^{2} \left(2 v - 1\right)}{256 \pi^{4}}\Big]\Big\}\\
c^{(0)}_{S_{11},d}\big|_{X\ln X}&=\left(\beta + 1\right)^{2} \Big[\frac{8 u w - 6 u + 8 w^{2} - 11 w + 3}{256 \pi^{4} w}\Big]
\end{align*}

\subsection{Gluon condensate in the quark-EMT correlator}
The contribution of the gluon condensate to the quark-EMT correlator contains logarithmic and pole terms in the representation~\eqref{eq:feynrep} and factorized terms,
\begin{equation}
\begin{split}
\Pi^{(d_4)}_{i,f}(p,p')={}&\int_{\Delta}du\,dw\sum_{k,l}c^{(d_4)}_{i,f;kl}(u,w)\,X^k(\ln X)^l\\
&+\sum c^{(d_4)}_{i,f}\,X_x^{k_x}(\ln X_x)^{l_x}X_y^{k_y}(\ln X_y)^{l_y},
\end{split}
\end{equation}
with $X_x=-p^2$ and $X_y=-p'^2$. The subscripts below label the corresponding functions.
\begin{align*}
c^{(d_4)}_{S_1,u}\big|_{X^{-1}}&=\left(\beta^{2} + 1\right) \Big[- \frac{\langle g^2G^2\rangle v}{1152 \pi^{4} w^{2}}\Big(\\
&\quad32 u^{4} + 96 u^{3} w - 46 u^{3} + 131 u^{2} w^{2} - 113 u^{2} w + 14 u^{2} + 102 u w^{3} - 95 u w^{2} + 26 u w\\
&\quad+35 w^{4} - 28 w^{3} + 8 w^{2}\Big)\Big] +\beta \Big[- \frac{\langle g^2G^2\rangle v}{384 \pi^{4} w^{2}}\Big(\\
&\quad8 u^{4} + 24 u^{3} w - 12 u^{3} + 48 u^{2} w^{2} - 30 u^{2} w + 4 u^{2} + 56 u w^{3} - 39 u w^{2} + 7 u w + 24 w^{4}\\
&\quad- 21 w^{3} + 4 w^{2}\Big)\Big]\\
c^{(d_4)}_{S_1,d}\big|_{X^{-1}}&=\left(\beta^{2} + 1\right) \Big[\frac{\langle g^2G^2\rangle v}{1536 \pi^{4} w^{2}}\Big(\\
&\quad16 u^{4} + 48 u^{3} w - 28 u^{3} + 86 u^{2} w^{2} - 52 u^{2} w + 12 u^{2} + 92 u w^{3} - 57 u w^{2} + 7 u w + 38 w^{4}\\
&\quad- 33 w^{3} + 2 w^{2}\Big)\Big] +\beta \Big[\frac{\langle g^2G^2\rangle v}{2304 \pi^{4} w^{2}}\Big(\\
&\quad32 u^{4} + 96 u^{3} w - 52 u^{3} + 62 u^{2} w^{2} - 92 u^{2} w + 20 u^{2} - 36 u w^{3} - 5 u w^{2} + 11 u w - 34 w^{4}\\
&\quad+35 w^{3} - 10 w^{2}\Big)\Big]\\
c^{(d_4)}_{S_2,u}\big|_{X^{-1}}&=\left(\beta^{2} + 1\right) \Big[- \frac{\langle g^2G^2\rangle}{2304 \pi^{4} w^{2}}\Big(\\
&\quad64 u^{5} + 192 u^{4} w - 164 u^{4} + 262 u^{3} w^{2} - 406 u^{3} w + 136 u^{3} + 204 u^{2} w^{3} - 442 u^{2} w^{2}\\
&\quad+241 u^{2} w - 36 u^{2} + 70 u w^{4} - 254 u w^{3} + 169 u w^{2} - 27 u w - 54 w^{4} + 45 w^{3} - 9 w^{2}\Big)\Big] +\beta\\
&\quad\Big[- \frac{\langle g^2G^2\rangle}{768 \pi^{4} w^{2}}\Big(\\
&\quad16 u^{5} + 48 u^{4} w - 40 u^{4} + 96 u^{3} w^{2} - 100 u^{3} w + 32 u^{3} + 112 u^{2} w^{3} - 146 u^{2} w^{2} + 59 u^{2} w\\
&\quad- 8 u^{2} + 48 u w^{4} - 104 u w^{3} + 53 u w^{2} - 7 u w - 18 w^{4} + 15 w^{3} - 3 w^{2}\Big)\Big]\\
c^{(d_4)}_{S_2,d}\big|_{X^{-1}}&=\left(\beta^{2} + 1\right) \Big[\frac{\langle g^2G^2\rangle}{3072 \pi^{4} w^{2}}\Big(\\
&\quad32 u^{5} + 96 u^{4} w - 72 u^{4} + 172 u^{3} w^{2} - 144 u^{3} w + 48 u^{3} + 184 u^{2} w^{3} - 158 u^{2} w^{2}\\
&\quad+53 u^{2} w - 8 u^{2} + 76 u w^{4} - 92 u w^{3} + 27 u w^{2} - 5 u w - 6 w^{4} + 5 w^{3} - w^{2}\Big)\Big] +\beta \Big[\frac{\langle g^2G^2\rangle}{4608 \pi^{4} w^{2}}\Big(\\
&\quad64 u^{5} + 192 u^{4} w - 152 u^{4} + 124 u^{3} w^{2} - 304 u^{3} w + 112 u^{3} - 72 u^{2} w^{3} - 70 u^{2} w^{2}\\
&\quad+121 u^{2} w - 24 u^{2} - 68 u w^{4} + 100 u w^{3} - 17 u w^{2} - 9 u w + 18 w^{4} - 15 w^{3} + 3 w^{2}\Big)\Big]\\
c^{(d_4)}_{S_5,u}\big|_{X^{-1}}&=\left(\beta^{2} + 1\right) \Big[- \frac{\langle g^2G^2\rangle}{4608 \pi^{4} w^{2}}\Big(\\
&\quad128 u^{5} + 384 u^{4} w - 312 u^{4} + 524 u^{3} w^{2} - 772 u^{3} w + 240 u^{3} + 408 u^{2} w^{3} - 780 u^{2} w^{2}\\
&\quad+491 u^{2} w - 56 u^{2} + 140 u w^{4} - 392 u w^{3} + 307 u w^{2} - 103 u w - 72 w^{4} + 69 w^{3} - 15 w^{2}\Big)\Big] +\beta\\
&\quad\Big[- \frac{\langle g^2G^2\rangle}{1536 \pi^{4} w^{2}}\Big( 32 u^{5} + 96 u^{4} w - 80 u^{4} + 192 u^{3} w^{2} - 200 u^{3} w + 64 u^{3} + 224 u^{2} w^{3} - 316 u^{2} w^{2}\\
&\quad+131 u^{2} w - 16 u^{2} + 96 u w^{4} - 244 u w^{3} + 143 u w^{2} - 27 u w - 48 w^{4} + 55 w^{3} - 13 w^{2}\Big)\Big]\\
c^{(d_4)}_{S_5,d}\big|_{X^{-1}}&=\left(\beta^{2} + 1\right) \Big[\frac{\langle g^2G^2\rangle}{6144 \pi^{4} w^{2}}\Big(\\
&\quad64 u^{5} + 192 u^{4} w - 176 u^{4} + 344 u^{3} w^{2} - 368 u^{3} w + 160 u^{3} + 368 u^{2} w^{3} - 500 u^{2} w^{2}\\
&\quad+185 u^{2} w - 48 u^{2} + 152 u w^{4} - 380 u w^{3} + 189 u w^{2} - 9 u w - 72 w^{4} + 93 w^{3} - 23 w^{2}\Big)\Big] +\beta\\
&\quad\Big[\frac{\langle g^2G^2\rangle}{9216 \pi^{4} w^{2}}\Big( 128 u^{5} + 384 u^{4} w - 336 u^{4} + 248 u^{3} w^{2} - 688 u^{3} w + 288 u^{3} - 144 u^{2} w^{3} - 132 u^{2} w^{2}\\
&\quad+317 u^{2} w - 80 u^{2} - 136 u w^{4} + 292 u w^{3} - 95 u w^{2} - 13 u w + 72 w^{4} - 87 w^{3} + 21 w^{2}\Big)\Big]\\
c^{(d_4)}_{S_7,u}\big|_{\ln X}&=\left(\beta^{2} + 1\right) \Big[\frac{\langle g^2G^2\rangle}{2304 \pi^{4} u w^{2}}\Big(\\
&\quad32 u^{4} + 64 u^{3} w - 46 u^{3} + 67 u^{2} w^{2} - 42 u^{2} w + 14 u^{2} + 35 u w^{3} - 4 u w^{2} - 4 u w + 6 w^{3} - 3 w^{2}\Big)\Big]\\
&\quad+\beta\\
&\quad\Big[\frac{\langle g^2G^2\rangle}{768 \pi^{4} u w^{2}}\left(8 u^{4} + 16 u^{3} w - 12 u^{3} + 32 u^{2} w^{2} - 12 u^{2} w + 4 u^{2} + 24 u w^{3} - 15 u w^{2} - 2 w^{3} + w^{2}\right)\Big]\\
c^{(d_4)}_{S_7,d}\big|_{\ln X}&=\left(\beta^{2} + 1\right) \Big[- \frac{\langle g^2G^2\rangle}{3072 \pi^{4} u w^{2}}\Big(\\
&\quad16 u^{4} + 32 u^{3} w - 28 u^{3} + 54 u^{2} w^{2} - 36 u^{2} w + 12 u^{2} + 38 u w^{3} - 43 u w^{2} + 8 u w - 10 w^{3} +5 w^{2}\Big)\Big]\\
&\quad+\beta \Big[- \frac{\langle g^2G^2\rangle}{4608 \pi^{4} u w^{2}}\Big(\\
&\quad32 u^{4} + 64 u^{3} w - 52 u^{3} - 2 u^{2} w^{2} - 60 u^{2} w + 20 u^{2} - 34 u w^{3} + 17 u w^{2} + 8 u w + 6 w^{3} - 3 w^{2}\Big)\Big]\\
c^{(d_4)}_{S_9,u}\big|_{\ln X}&=\left(\beta^{2} + 1\right) \Big[- \frac{\langle g^2G^2\rangle}{9216 \pi^{4} u v w^{2}}\Big(\\
&\quad128 u^{5} + 384 u^{4} w - 320 u^{4} + 524 u^{3} w^{2} - 900 u^{3} w + 256 u^{3} + 408 u^{2} w^{3} - 1048 u^{2} w^{2}\\
&\quad+686 u^{2} w - 64 u^{2} + 140 u w^{4} - 576 u w^{3} + 570 u w^{2} - 170 u w - 108 w^{4} + 153 w^{3} - 45 w^{2}\Big)\Big] +\beta\\
&\quad\Big[- \frac{\langle g^2G^2\rangle}{3072 \pi^{4} u v w^{2}}\Big( 32 u^{5} + 96 u^{4} w - 80 u^{4} + 192 u^{3} w^{2} - 228 u^{3} w + 64 u^{3} + 224 u^{2} w^{3} - 360 u^{2} w^{2}\\
&\quad+174 u^{2} w - 16 u^{2} + 96 u w^{4} - 248 u w^{3} + 182 u w^{2} - 42 u w - 36 w^{4} + 51 w^{3} - 15 w^{2}\Big)\Big]\\
c^{(d_4)}_{S_9,u}\big|_{\ln X_x\ln X_y}&=\left(\beta^{2} + 1\right) \Big[- \frac{\langle g^2G^2\rangle}{512 \pi^{4}}\Big] +\beta \Big[- \frac{\langle g^2G^2\rangle}{512 \pi^{4}}\Big]\\
c^{(d_4)}_{S_9,d}\big|_{\ln X}&=\left(\beta^{2} + 1\right) \Big[\frac{\langle g^2G^2\rangle}{12288 \pi^{4} u v w^{2}}\Big(\\
&\quad64 u^{5} + 192 u^{4} w - 160 u^{4} + 344 u^{3} w^{2} - 348 u^{3} w + 128 u^{3} + 368 u^{2} w^{3} - 448 u^{2} w^{2}\\
&\quad+170 u^{2} w - 32 u^{2} + 152 u w^{4} - 272 u w^{3} + 130 u w^{2} - 14 u w - 12 w^{4} + 17 w^{3} - 5 w^{2}\Big)\Big] +\beta\\
&\quad\Big[\frac{\langle g^2G^2\rangle}{18432 \pi^{4} u v w^{2}}\Big(\\
&\quad128 u^{5} + 384 u^{4} w - 320 u^{4} + 248 u^{3} w^{2} - 684 u^{3} w + 256 u^{3} - 144 u^{2} w^{3} - 208 u^{2} w^{2}\\
&\quad+338 u^{2} w - 64 u^{2} - 136 u w^{4} + 192 u w^{3} - 6 u w^{2} - 38 u w + 36 w^{4} - 51 w^{3} + 15 w^{2}\Big)\Big]\\
c^{(d_4)}_{S_9,d}\big|_{\ln X_x\ln X_y}&=\left(1 - \beta\right)^{2} \Big[\frac{\langle g^2G^2\rangle}{6144 \pi^{4}}\Big]\\
c^{(d_4)}_{S_{11},u}\big|_{X^{-1}}&=\left(\beta^{2} + 1\right) \Big[- \frac{\langle g^2G^2\rangle \left(2 v - 1\right)}{3072 \pi^{4} w}\left(20 u^{2} + 20 u w - 20 u + 21 w^{2} - 7 w\right)\Big] +\beta\\
&\quad\Big[- \frac{\langle g^2G^2\rangle \left(2 v - 1\right) \left(4 u^{2} + 4 u w - 4 u + 3 w^{2} - w\right)}{768 \pi^{4} w}\Big]\\
c^{(d_4)}_{S_{11},d}\big|_{X^{-1}}&=\left(\beta^{2} + 1\right) \Big[\frac{\langle g^2G^2\rangle \left(2 v - 1\right) \left(20 u^{2} + 20 u w - 20 u - 3 w^{2} + w\right)}{6144 \pi^{4} w}\Big] +\beta\\
&\quad\Big[\frac{\langle g^2G^2\rangle \left(2 v - 1\right) \left(12 u^{2} + 12 u w - 12 u + 3 w^{2} - w\right)}{3072 \pi^{4} w}\Big]
\end{align*}

\subsection{Four-quark and mixed condensates}
The contributions of the four-quark condensate and of the mixed condensate consist of factorized terms only,
\begin{equation}
\Pi^{(d_6,d_8)}_{i,f}(p,p')=\sum c^{(d_6,d_8)}_{i,f}\,X_x^{k_x}X_y^{k_y}.
\end{equation}
All of them are proportional to $\beta-1$ and vanish at $\beta=1$.
\begin{align*}
c^{(d_6)}_{S_2,u}\big|_{X_x^{-1}X_y^{-1}}&=\Big[- \frac{\langle\bar qq\rangle^{2}}{2}\Big] +\beta^{2} \Big[\frac{\langle\bar qq\rangle^{2}}{2}\Big]\\
c^{(d_6)}_{S_2,d}\big|_{X_x^{-1}X_y^{-1}}&=\left(1 - \beta\right)^{2} \Big[\frac{\langle\bar qq\rangle^{2}}{12}\Big]\\
c^{(d_6)}_{S_5,u}\big|_{X_x^{-1}X_y^{-1}}&=\Big[- \frac{\langle\bar qq\rangle^{2}}{4}\Big] +\beta^{2} \Big[\frac{\langle\bar qq\rangle^{2}}{4}\Big]\\
c^{(d_6)}_{S_5,d}\big|_{X_x^{-1}X_y^{-1}}&=\left(1 - \beta\right)^{2} \Big[\frac{\langle\bar qq\rangle^{2}}{24}\Big]\\
c^{(d_6)}_{S_9,u}\big|_{X_x^{-1}X_y^{-1}}&=\Big\{ (p\cdot p^{\prime})\,\Big[\frac{\langle\bar qq\rangle^{2}}{4}\Big]\Big\} +\beta^{2}\Big\{ (p\cdot p^{\prime})\,\Big[- \frac{\langle\bar qq\rangle^{2}}{4}\Big]\Big\}\\
c^{(d_6)}_{S_9,d}\big|_{X_x^{-1}X_y^{-1}}&=\left(1 - \beta\right)^{2}\Big\{ (p\cdot p^{\prime})\,\Big[- \frac{\langle\bar qq\rangle^{2}}{24}\Big]\Big\}\\
c^{(d_6)}_{S_{11},u}\big|_{X_x^{-1}X_y^{-1}}&=\Big[- \frac{\langle\bar qq\rangle^{2}}{4}\Big] +\beta^{2} \Big[\frac{\langle\bar qq\rangle^{2}}{4}\Big]\\
c^{(d_6)}_{S_{11},d}\big|_{X_x^{-1}X_y^{-1}}&=\left(1 - \beta\right)^{2} \Big[- \frac{\langle\bar qq\rangle^{2}}{24}\Big]
\end{align*}
\begin{align*}
c^{(d_8)}_{S_1,u}\big|_{X_x^{-2}X_y^{-1}}&=\Big[\frac{m_0^2 \langle\bar qq\rangle^{2}}{2}\Big] +\beta^{2} \Big[- \frac{m_0^2 \langle\bar qq\rangle^{2}}{2}\Big]\\
c^{(d_8)}_{S_1,d}\big|_{X_x^{-2}X_y^{-1}}&=\left(1 - \beta\right)^{2} \Big[- \frac{m_0^2 \langle\bar qq\rangle^{2}}{12}\Big]\\
c^{(d_8)}_{S_2,u}\big|_{X_x^{-1}X_y^{-2}}&=\Big[\frac{m_0^2 \langle\bar qq\rangle^{2}}{4}\Big] +\beta^{2} \Big[- \frac{m_0^2 \langle\bar qq\rangle^{2}}{4}\Big]\\
c^{(d_8)}_{S_2,u}\big|_{X_x^{-2}X_y^{-1}}&=\Big[\frac{3 m_0^2 \langle\bar qq\rangle^{2}}{4}\Big] +\beta^{2} \Big[- \frac{3 m_0^2 \langle\bar qq\rangle^{2}}{4}\Big]\\
c^{(d_8)}_{S_2,u}\big|_{X_x^{-2}X_y^{-2}}&=\Big\{ (p\cdot p^{\prime})\,\Big[\frac{m_0^2 \langle\bar qq\rangle^{2}}{2}\Big]\Big\} +\beta^{2}\Big\{ (p\cdot p^{\prime})\,\Big[- \frac{m_0^2 \langle\bar qq\rangle^{2}}{2}\Big]\Big\}\\
c^{(d_8)}_{S_2,d}\big|_{X_x^{-1}X_y^{-2}}&=\left(1 - \beta\right)^{2} \Big[- \frac{m_0^2 \langle\bar qq\rangle^{2}}{48}\Big]\\
c^{(d_8)}_{S_2,d}\big|_{X_x^{-2}X_y^{-1}}&=\left(1 - \beta\right)^{2} \Big[- \frac{19 m_0^2 \langle\bar qq\rangle^{2}}{144}\Big]\\
c^{(d_8)}_{S_2,d}\big|_{X_x^{-2}X_y^{-2}}&=\left(1 - \beta\right)^{2}\Big\{ (p\cdot p^{\prime})\,\Big[- \frac{7 m_0^2 \langle\bar qq\rangle^{2}}{72}\Big]\Big\}\\
c^{(d_8)}_{S_5,u}\big|_{X_x^{-1}X_y^{-2}}&=\Big[\frac{m_0^2 \langle\bar qq\rangle^{2}}{8}\Big] +\beta^{2} \Big[- \frac{m_0^2 \langle\bar qq\rangle^{2}}{8}\Big]\\
c^{(d_8)}_{S_5,u}\big|_{X_x^{-2}X_y^{-1}}&=\Big[\frac{5 m_0^2 \langle\bar qq\rangle^{2}}{8}\Big] +\beta^{2} \Big[- \frac{5 m_0^2 \langle\bar qq\rangle^{2}}{8}\Big]\\
c^{(d_8)}_{S_5,u}\big|_{X_x^{-2}X_y^{-2}}&=\Big\{ (p\cdot p^{\prime})\,\Big[\frac{m_0^2 \langle\bar qq\rangle^{2}}{4}\Big]\Big\} +\beta^{2}\Big\{ (p\cdot p^{\prime})\,\Big[- \frac{m_0^2 \langle\bar qq\rangle^{2}}{4}\Big]\Big\}\\
c^{(d_8)}_{S_5,d}\big|_{X_x^{-1}X_y^{-2}}&=\left(1 - \beta\right)^{2} \Big[- \frac{5 m_0^2 \langle\bar qq\rangle^{2}}{288}\Big]\\
c^{(d_8)}_{S_5,d}\big|_{X_x^{-2}X_y^{-1}}&=\left(1 - \beta\right)^{2} \Big[- \frac{25 m_0^2 \langle\bar qq\rangle^{2}}{288}\Big]\\
c^{(d_8)}_{S_5,d}\big|_{X_x^{-2}X_y^{-2}}&=\left(1 - \beta\right)^{2}\Big\{ (p\cdot p^{\prime})\,\Big[- \frac{7 m_0^2 \langle\bar qq\rangle^{2}}{144}\Big]\Big\}\\
c^{(d_8)}_{S_7,u}\big|_{X_x^{-1}X_y^{-1}}&=\Big[\frac{m_0^2 \langle\bar qq\rangle^{2}}{4}\Big] +\beta^{2} \Big[- \frac{m_0^2 \langle\bar qq\rangle^{2}}{4}\Big]\\
c^{(d_8)}_{S_7,d}\big|_{X_x^{-1}X_y^{-1}}&=\left(1 - \beta\right)^{2} \Big[- \frac{m_0^2 \langle\bar qq\rangle^{2}}{24}\Big]\\
c^{(d_8)}_{S_7,d}\big|_{X_x^{-2}X_y^{-1}}&=\left(1 - \beta\right)^{2}\Big\{ (p\cdot p^{\prime})\,\Big[- \frac{m_0^2 \langle\bar qq\rangle^{2}}{72}\Big]\Big\}\\
c^{(d_8)}_{S_9,u}\big|_{X_x^{-1}X_y^{-1}}&=\Big[- \frac{m_0^2 \langle\bar qq\rangle^{2}}{8}\Big] +\beta^{2} \Big[\frac{m_0^2 \langle\bar qq\rangle^{2}}{8}\Big]\\
c^{(d_8)}_{S_9,u}\big|_{X_x^{-1}X_y^{-2}}&=\Big\{ (p\cdot p^{\prime})\,\Big[- \frac{m_0^2 \langle\bar qq\rangle^{2}}{8}\Big] +p^{\prime 2}\,\Big[- \frac{m_0^2 \langle\bar qq\rangle^{2}}{8}\Big]\Big\} +\beta^{2}\Big\{ (p\cdot p^{\prime})\,\Big[\frac{m_0^2 \langle\bar qq\rangle^{2}}{8}\Big]\\
&\quad+p^{\prime 2}\,\Big[\frac{m_0^2 \langle\bar qq\rangle^{2}}{8}\Big]\Big\}\\
c^{(d_8)}_{S_9,u}\big|_{X_x^{-2}X_y^{-1}}&=\Big\{ (p\cdot p^{\prime})\,\Big[- \frac{3 m_0^2 \langle\bar qq\rangle^{2}}{8}\Big] +p^{2}\,\Big[- \frac{m_0^2 \langle\bar qq\rangle^{2}}{8}\Big]\Big\} +\beta^{2}\Big\{ (p\cdot p^{\prime})\,\Big[\frac{3 m_0^2 \langle\bar qq\rangle^{2}}{8}\Big]\\
&\quad+p^{2}\,\Big[\frac{m_0^2 \langle\bar qq\rangle^{2}}{8}\Big]\Big\}\\
c^{(d_8)}_{S_9,u}\big|_{X_x^{-2}X_y^{-2}}&=\Big\{ (p\cdot p^{\prime})^{2}\,\Big[- \frac{m_0^2 \langle\bar qq\rangle^{2}}{4}\Big]\Big\} +\beta^{2}\Big\{ (p\cdot p^{\prime})^{2}\,\Big[\frac{m_0^2 \langle\bar qq\rangle^{2}}{4}\Big]\Big\}\\
c^{(d_8)}_{S_9,d}\big|_{X_x^{-1}X_y^{-1}}&=\left(1 - \beta\right)^{2} \Big[\frac{7 m_0^2 \langle\bar qq\rangle^{2}}{288}\Big]\\
c^{(d_8)}_{S_9,d}\big|_{X_x^{-1}X_y^{-2}}&=\left(1 - \beta\right)^{2}\Big\{ (p\cdot p^{\prime})\,\Big[\frac{5 m_0^2 \langle\bar qq\rangle^{2}}{288}\Big] +p^{\prime 2}\,\Big[\frac{7 m_0^2 \langle\bar qq\rangle^{2}}{288}\Big]\Big\}\\
c^{(d_8)}_{S_9,d}\big|_{X_x^{-2}X_y^{-1}}&=\left(1 - \beta\right)^{2}\Big\{ (p\cdot p^{\prime})\,\Big[\frac{19 m_0^2 \langle\bar qq\rangle^{2}}{288}\Big] +p^{2}\,\Big[\frac{7 m_0^2 \langle\bar qq\rangle^{2}}{288}\Big]\Big\}\\
c^{(d_8)}_{S_9,d}\big|_{X_x^{-2}X_y^{-2}}&=\left(1 - \beta\right)^{2}\Big\{ (p\cdot p^{\prime})^{2}\,\Big[\frac{7 m_0^2 \langle\bar qq\rangle^{2}}{144}\Big]\Big\}\\
c^{(d_8)}_{S_{11},u}\big|_{X_x^{-1}X_y^{-2}}&=\Big[\frac{m_0^2 \langle\bar qq\rangle^{2}}{8}\Big] +\beta^{2} \Big[- \frac{m_0^2 \langle\bar qq\rangle^{2}}{8}\Big]\\
c^{(d_8)}_{S_{11},u}\big|_{X_x^{-2}X_y^{-1}}&=\Big[\frac{3 m_0^2 \langle\bar qq\rangle^{2}}{8}\Big] +\beta^{2} \Big[- \frac{3 m_0^2 \langle\bar qq\rangle^{2}}{8}\Big]\\
c^{(d_8)}_{S_{11},u}\big|_{X_x^{-2}X_y^{-2}}&=\Big\{ (p\cdot p^{\prime})\,\Big[\frac{m_0^2 \langle\bar qq\rangle^{2}}{4}\Big]\Big\} +\beta^{2}\Big\{ (p\cdot p^{\prime})\,\Big[- \frac{m_0^2 \langle\bar qq\rangle^{2}}{4}\Big]\Big\}\\
c^{(d_8)}_{S_{11},d}\big|_{X_x^{-1}X_y^{-2}}&=\left(1 - \beta\right)^{2} \Big[\frac{5 m_0^2 \langle\bar qq\rangle^{2}}{288}\Big]\\
c^{(d_8)}_{S_{11},d}\big|_{X_x^{-2}X_y^{-1}}&=\left(1 - \beta\right)^{2} \Big[\frac{19 m_0^2 \langle\bar qq\rangle^{2}}{288}\Big]\\
c^{(d_8)}_{S_{11},d}\big|_{X_x^{-2}X_y^{-2}}&=\left(1 - \beta\right)^{2}\Big\{ (p\cdot p^{\prime})\,\Big[\frac{7 m_0^2 \langle\bar qq\rangle^{2}}{144}\Big]\Big\}
\end{align*}

\subsection{Gluon-EMT correlator}
At LO the gluon-EMT correlator consists of condensate terms only. The gluon-condensate (dimension-four) part reads
\begin{equation}
\Pi^{(g_4)}_{i}(p,p')=-\frac{1}{Q^2}\int_{\Delta}du\,dw\;\sum_{k}c^{(g_4)}_{i;k}(u,w)\,X^k\ln X ,
\end{equation}
and the dimension-eight part is $\Pi^{(g_8)}_i=-(1/Q^2)\sum c^{(g_8)}_{i}\,X_x^{k_x}X_y^{k_y}$, which is proportional to $\beta-1$. The coefficients are summed over the attachments to the $u$- and $d$-quark lines.
\begin{align*}
c^{(g_4)}_{S_1}\big|_{\ln X}&=\left(\beta^{2} + 1\right)\\
&\quad\Big[- \frac{5 \langle g^2G^2\rangle v}{1152 \pi^{4} u w^{2}}\left(16 u^{3} w + 4 u^{3} + 32 u^{2} w^{2} - 3 u^{2} w - u^{2} + 14 u w^{3} - 11 u w^{2} - 2 w^{4} - 2 w^{3} + w^{2}\right)\Big]\\
&\quad+\beta\\
&\quad\Big[- \frac{\langle g^2G^2\rangle v}{576 \pi^{4} u w^{2}}\left(16 u^{3} w + 4 u^{3} + 32 u^{2} w^{2} - 3 u^{2} w - u^{2} + 14 u w^{3} - 11 u w^{2} - 2 w^{4} - 2 w^{3} + w^{2}\right)\Big]\\
c^{(g_4)}_{S_2}\big|_{\ln X}&=\left(\beta^{2} + 1\right) \Big[- \frac{5 \langle g^2G^2\rangle}{1152 \pi^{4} w^{2}}\Big(\\
&\quad16 u^{3} w + 4 u^{3} + 32 u^{2} w^{2} - 21 u^{2} w - 7 u^{2} + 14 u w^{3} - 38 u w^{2} + 6 u w + 3 u - 2 w^{4}\\
&\quad- 11 w^{3} + 7 w^{2}\Big)\Big] +\beta \Big[- \frac{\langle g^2G^2\rangle}{576 \pi^{4} w^{2}}\Big(\\
&\quad16 u^{3} w + 4 u^{3} + 32 u^{2} w^{2} - 21 u^{2} w - 7 u^{2} + 14 u w^{3} - 38 u w^{2} + 6 u w + 3 u - 2 w^{4}\\
&\quad- 11 w^{3} + 7 w^{2}\Big)\Big]\\
c^{(g_4)}_{S_5}\big|_{\ln X}&=\left(\beta^{2} + 1\right) \Big[- \frac{5 \langle g^2G^2\rangle}{2304 \pi^{4} w^{2}}\Big(\\
&\quad32 u^{3} w + 8 u^{3} + 64 u^{2} w^{2} - 30 u^{2} w - 10 u^{2} + 28 u w^{3} - 58 u w^{2} + 4 u w + 2 u - 4 w^{4}\\
&\quad- 16 w^{3} + 17 w^{2}\Big)\Big] +\beta \Big[- \frac{\langle g^2G^2\rangle}{1152 \pi^{4} w^{2}}\Big(\\
&\quad32 u^{3} w + 8 u^{3} + 64 u^{2} w^{2} - 30 u^{2} w - 10 u^{2} + 28 u w^{3} - 58 u w^{2} + 4 u w + 2 u - 4 w^{4}\\
&\quad- 16 w^{3} + 17 w^{2}\Big)\Big]\\
c^{(g_4)}_{S_7}\big|_{X\ln X}&=\left(\beta^{2} + 1\right) \Big[\frac{5 \langle g^2G^2\rangle}{2304 \pi^{4} u w}\left(16 u^{2} + 16 u w - 7 u - 2 w^{2} + 7 w - 2\right)\Big] +\beta\\
&\quad\Big[\frac{\langle g^2G^2\rangle}{1152 \pi^{4} u w}\left(16 u^{2} + 16 u w - 7 u - 2 w^{2} + 7 w - 2\right)\Big]\\
c^{(g_4)}_{S_9}\big|_{X\ln X}&=\left(\beta^{2} + 1\right)\\
&\quad\Big[- \frac{5 \langle g^2G^2\rangle}{4608 \pi^{4} u v w}\left(32 u^{3} + 64 u^{2} w - 52 u^{2} + 28 u w^{2} - 92 u w + 34 u - 4 w^{3} - 36 w^{2} + 48 w - 11\right)\Big]\\
&\quad+\beta\\
&\quad\Big[- \frac{\langle g^2G^2\rangle}{2304 \pi^{4} u v w}\left(32 u^{3} + 64 u^{2} w - 52 u^{2} + 28 u w^{2} - 92 u w + 34 u - 4 w^{3} - 36 w^{2} + 48 w - 11\right)\Big]\\
c^{(g_4)}_{S_{11}}\big|_{\ln X}&=\left(\beta^{2} + 1\right) \Big[\frac{5 \langle g^2G^2\rangle \left(3 u + 3 w - 2\right)}{768 \pi^{4}}\Big] +\beta \Big[\frac{\langle g^2G^2\rangle \left(3 u + 3 w - 2\right)}{384 \pi^{4}}\Big]
\end{align*}
\begin{align*}
c^{(g_8)}_{S_2}\big|_{X_x^{-1}X_y^{-1}}&=\left(1 - \beta\right)^{2} \Big[- \frac{m_0^2 \langle\bar qq\rangle^{2}}{36}\Big]\\
c^{(g_8)}_{S_5}\big|_{X_x^{-1}X_y^{-1}}&=\left(1 - \beta\right)^{2} \Big[\frac{m_0^2 \langle\bar qq\rangle^{2}}{72}\Big]\\
c^{(g_8)}_{S_9}\big|_{X_x^{-1}X_y^{-1}}&=\left(1 - \beta\right)^{2}\Big\{ (p\cdot p^{\prime})\,\Big[- \frac{m_0^2 \langle\bar qq\rangle^{2}}{72}\Big]\Big\}\\
c^{(g_8)}_{S_{11}}\big|_{X_x^{-1}X_y^{-1}}&=\left(1 - \beta\right)^{2} \Big[- \frac{m_0^2 \langle\bar qq\rangle^{2}}{72}\Big]
\end{align*}

The normalization of the coefficients corresponds to $\lambda^2e^{-m^2(\tau_1+\tau_2)}$ times the pole coefficients of Table~\ref{tab:pole} on the hadronic side.

\subsection{\texorpdfstring{$\mathcal{O}(\alpha_s)$}{O(alpha s)} subtraction terms}

The $\mathcal{O}(\alpha_s)$ contributions are computed numerically and are not listed here. The divergent subgraphs are subtracted locally and added back with their $D$-dimensional $\overline{\rm MS}$ finite parts. For the local EMT vertex functions the finite parts differ from those with the Dirac algebra fixed to four dimensions by $-2/3$ (quark EMT) and $-8/3$ (gluon EMT), in units of $\alpha_sC_F/(4\pi)$ times the LO vertex, in the structure $\tfrac12(q_\mu\gamma_\nu+q_\nu\gamma_\mu)$. There is no difference in the structure $q_\mu q_\nu\slashed q/q^2$, and the trace-type component proportional to $g_{\mu\nu}\slashed q$ yields a contact term that vanishes under the double Borel transformation. The subgraphs of the interpolating currents are subtracted in the structure-preserving prescription of Sec.~\ref{sec:ope}.

The numerical tables of the $\mathcal{O}(\alpha_s)$ contributions are available from the author on request. As a guide, Table~\ref{tab:oasratio} gives the ratios of the $\mathcal{O}(\alpha_s)$ contributions of the quark and gluon EMT to the LO perturbative contribution, for the protected combinations of $A$, $J$ and $D$ at a point inside the working region.

\begin{table}[htb]
\caption{Ratios $r_q$ and $r_g$ of the $\mathcal{O}(\alpha_s)$ contributions of the quark EMT ($u+d$) and of the gluon EMT to the LO perturbative contribution, for the protected combinations $A=H_1+H_2+2H_5$, $J=2H_{11}$ and $D=\tfrac12(D^{(1)}+D^{(3)})$, at $\beta=1$, $M^2=0.80$~GeV$^2$, $S_0=4.0$~GeV$^2$ and $\mu=1$~GeV, with $\alpha_s(1\,{\rm GeV})=0.47$. The point lies inside the working region of all three form factors.}
\label{tab:oasratio}
\renewcommand{\arraystretch}{1.3}
\begin{ruledtabular}
\begin{tabular}{ccccccc}
 & \multicolumn{2}{c}{$A$} & \multicolumn{2}{c}{$J$} & \multicolumn{2}{c}{$D$}\\
$Q^2$ (GeV$^2$) & $r_q$ & $r_g$ & $r_q$ & $r_g$ & $r_q$ & $r_g$\\
\hline
1 & $1.025$ & $0.189$ & $1.027$ & $0.178$ & $0.837$ & $0.373$\\
1.5 & $1.087$ & $0.161$ & $1.080$ & $0.156$ & $0.911$ & $0.369$\\
2 & $1.112$ & $0.136$ & $1.113$ & $0.128$ & $0.911$ & $0.347$\\
2.5 & $1.146$ & $0.113$ & $1.143$ & $0.107$ & $0.970$ & $0.299$\\
\end{tabular}
\end{ruledtabular}
\end{table}

\section{Numerical uncertainty of the \texorpdfstring{$\mathcal{O}(\alpha_s)$}{O(alpha s)} integrals}\label{app:numerics}
The numerical uncertainty has four contributions. (i) The statistical error of each part, propagated with all kinematic points shifted coherently and combined in quadrature over parts. The symmetric and antisymmetric productions are independent runs. (ii) The reproducibility, estimated from independent repetitions at each kinematic point, applied coherently within the quark or gluon contribution and in quadrature between flavours. For the antisymmetric moments, the scale is $|\mathrm{LO}_u|+|\mathrm{LO}_d|$. (iii) The convergence with the integration budget, from a dedicated audit at $x=1.25$ with twice (four times for the two slowest-converging parts) the budget. For the antisymmetric moments, the relative convergence of the corresponding symmetric class is applied (an assumption). (iv) The interpolation in $x$, estimated by replacing the cubic interpolation with a monotone one. Its effect on the form factors is of order $10^{-5}$.

\bibliographystyle{elsarticle-num}
\bibliography{gff_oas_refs}

@article{Kobzarev:1962wt,
    author = "Kobzarev, I. Yu. and Okun', L. B.",
    title = "{Gravitational Interaction of Fermions}",
    journal = "Zh. Eksp. Teor. Fiz.",
    volume = "43",
    pages = "1904--1909",
    year = "1962"
}

@article{Pagels:1966zza,
    author = "Pagels, Heinz",
    title = "{Energy-Momentum Structure Form Factors of Particles}",
    doi = "10.1103/PhysRev.144.1250",
    journal = "Phys. Rev.",
    volume = "144",
    pages = "1250--1260",
    year = "1966"
}

@article{Muller:1994ses,
    author = {M{\"u}ller, Dieter and Robaschik, D. and Geyer, B. and Dittes, F. -M. and Ho{\v{r}}ej{\v{s}}i, J.},
    title = "{Wave functions, evolution equations and evolution kernels from light ray operators of QCD}",
    eprint = "hep-ph/9812448",
    archivePrefix = "arXiv",
    reportNumber = "NTZ-6-91, NTZ-91-6",
    doi = "10.1002/prop.2190420202",
    journal = "Fortsch. Phys.",
    volume = "42",
    pages = "101--141",
    year = "1994"
}

@article{Ji:1996ek,
    author = "Ji, Xiang-Dong",
    title = "{Gauge-Invariant Decomposition of Nucleon Spin}",
    eprint = "hep-ph/9603249",
    archivePrefix = "arXiv",
    reportNumber = "MIT-CTP-2517",
    doi = "10.1103/PhysRevLett.78.610",
    journal = "Phys. Rev. Lett.",
    volume = "78",
    pages = "610--613",
    year = "1997"
}

@article{Radyushkin:1996nd,
    author = "Radyushkin, A. V.",
    title = "{Scaling limit of deeply virtual Compton scattering}",
    eprint = "hep-ph/9604317",
    archivePrefix = "arXiv",
    reportNumber = "CEBAF-TH-96-05",
    doi = "10.1016/0370-2693(96)00528-X",
    journal = "Phys. Lett. B",
    volume = "380",
    pages = "417--425",
    year = "1996"
}

@article{Goeke:2001tz,
    author = "Goeke, K. and Polyakov, Maxim V. and Vanderhaeghen, M.",
    title = "{Hard exclusive reactions and the structure of hadrons}",
    eprint = "hep-ph/0106012",
    archivePrefix = "arXiv",
    doi = "10.1016/S0146-6410(01)00158-2",
    journal = "Prog. Part. Nucl. Phys.",
    volume = "47",
    pages = "401--515",
    year = "2001"
}

@article{Diehl:2003ny,
    author = "Diehl, M.",
    title = "{Generalized parton distributions}",
    eprint = "hep-ph/0307382",
    archivePrefix = "arXiv",
    reportNumber = "DESY-THESIS-2003-018",
    doi = "10.1016/j.physrep.2003.08.002",
    journal = "Phys. Rept.",
    volume = "388",
    pages = "41--277",
    year = "2003"
}

@article{Leader:2013jra,
    author = "Leader, E. and Lorc{\'e}, C.",
    title = "{The angular momentum controversy: What{\textquoteright}s it all about and does it matter?}",
    eprint = "1309.4235",
    archivePrefix = "arXiv",
    primaryClass = "hep-ph",
    doi = "10.1016/j.physrep.2014.02.010",
    journal = "Phys. Rept.",
    volume = "541",
    number = "3",
    pages = "163--248",
    year = "2014"
}

@article{Ji:1994av,
    author = "Ji, Xiang-Dong",
    title = "{A QCD analysis of the mass structure of the nucleon}",
    eprint = "hep-ph/9410274",
    archivePrefix = "arXiv",
    reportNumber = "MIT-CTP-2368",
    doi = "10.1103/PhysRevLett.74.1071",
    journal = "Phys. Rev. Lett.",
    volume = "74",
    pages = "1071--1074",
    year = "1995"
}

@article{Lorce:2017xzd,
    author = "Lorc{\'e}, C{\'e}dric",
    title = "{On the hadron mass decomposition}",
    eprint = "1706.05853",
    archivePrefix = "arXiv",
    primaryClass = "hep-ph",
    doi = "10.1140/epjc/s10052-018-5561-2",
    journal = "Eur. Phys. J. C",
    volume = "78",
    number = "2",
    pages = "120",
    year = "2018"
}

@article{Metz:2020vxd,
    author = "Metz, Andreas and Pasquini, Barbara and Rodini, Simone",
    title = "{Revisiting the proton mass decomposition}",
    eprint = "2006.11171",
    archivePrefix = "arXiv",
    primaryClass = "hep-ph",
    doi = "10.1103/PhysRevD.102.114042",
    journal = "Phys. Rev. D",
    volume = "102",
    pages = "114042",
    year = "2020"
}

@article{Lowdon:2017idv,
    author = "Lowdon, Peter and Chiu, Kelly Yu-Ju and Brodsky, Stanley J.",
    title = "{Rigorous constraints on the matrix elements of the energy-momentum tensor}",
    eprint = "1707.06313",
    archivePrefix = "arXiv",
    primaryClass = "hep-th",
    reportNumber = "SLAC-PUB-17111",
    doi = "10.1016/j.physletb.2017.09.050",
    journal = "Phys. Lett. B",
    volume = "774",
    pages = "1--6",
    year = "2017"
}

@article{Polyakov:1999gs,
    author = "Polyakov, Maxim V. and Weiss, C.",
    title = "{Skewed and double distributions in pion and nucleon}",
    eprint = "hep-ph/9902451",
    archivePrefix = "arXiv",
    reportNumber = "RUB-TPII-1-99",
    doi = "10.1103/PhysRevD.60.114017",
    journal = "Phys. Rev. D",
    volume = "60",
    pages = "114017",
    year = "1999"
}

@article{Polyakov:2002yz,
    author = "Polyakov, M. V.",
    title = "{Generalized parton distributions and strong forces inside nucleons and nuclei}",
    eprint = "hep-ph/0210165",
    archivePrefix = "arXiv",
    reportNumber = "RUB-TP2-14-02",
    doi = "10.1016/S0370-2693(03)00036-4",
    journal = "Phys. Lett. B",
    volume = "555",
    pages = "57--62",
    year = "2003"
}

@article{Polyakov:2018zvc,
    author = "Polyakov, Maxim V. and Schweitzer, Peter",
    title = "{Forces inside hadrons: pressure, surface tension, mechanical radius, and all that}",
    eprint = "1805.06596",
    archivePrefix = "arXiv",
    primaryClass = "hep-ph",
    doi = "10.1142/S0217751X18300259",
    journal = "Int. J. Mod. Phys. A",
    volume = "33",
    number = "26",
    pages = "1830025",
    year = "2018"
}

@article{Freese:2021czn,
    author = "Freese, Adam and Miller, Gerald A.",
    title = "{Forces within hadrons on the light front}",
    eprint = "2102.01683",
    archivePrefix = "arXiv",
    primaryClass = "hep-ph",
    reportNumber = "NT@UW-21-02",
    doi = "10.1103/PhysRevD.103.094023",
    journal = "Phys. Rev. D",
    volume = "103",
    pages = "094023",
    year = "2021"
}

@article{Lorce:2022cle,
    author = "Lorc{\'e}, C{\'e}dric and Schweitzer, Peter and Tezgin, Kemal",
    title = "{2D energy-momentum tensor distributions of nucleon in a large-Nc quark model from ultrarelativistic to nonrelativistic limit}",
    eprint = "2202.01192",
    archivePrefix = "arXiv",
    primaryClass = "hep-ph",
    doi = "10.1103/PhysRevD.106.014012",
    journal = "Phys. Rev. D",
    volume = "106",
    number = "1",
    pages = "014012",
    year = "2022"
}

@article{Ji:2021mfb,
    author = "Ji, Xiangdong and Liu, Yizhuang",
    title = "{Momentum-Current Gravitational Multipoles of Hadrons}",
    eprint = "2110.14781",
    archivePrefix = "arXiv",
    primaryClass = "hep-ph",
    doi = "10.1103/PhysRevD.106.034028",
    journal = "Phys. Rev. D",
    volume = "106",
    number = "3",
    pages = "034028",
    year = "2022"
}

@article{Freese:2021qtb,
    author = "Freese, Adam and Miller, Gerald A.",
    title = "{Genuine empirical pressure within the proton}",
    eprint = "2104.03213",
    archivePrefix = "arXiv",
    primaryClass = "hep-ph",
    reportNumber = "NT@UW-21-03",
    doi = "10.1103/PhysRevD.104.014024",
    journal = "Phys. Rev. D",
    volume = "104",
    number = "1",
    pages = "014024",
    year = "2021"
}

@article{Lorce:2025oot,
    author = "Lorc{\'e}, C{\'e}dric and Schweitzer, Peter",
    title = "{Pressure Inside Hadrons: Criticism, Conjectures, and All That}",
    eprint = "2501.04622",
    archivePrefix = "arXiv",
    primaryClass = "hep-ph",
    doi = "10.5506/aphyspolb.56.3-a17",
    journal = "Acta Phys. Polon. B",
    volume = "56",
    number = "3",
    pages = "3\textendash{}A17",
    year = "2025"
}

@article{Burkert:2023wzr,
    author = "Burkert, V. D. and Elouadrhiri, L. and Girod, F. X. and Lorc{\'e}, C. and Schweitzer, P. and Shanahan, P. E.",
    title = "{Colloquium: Gravitational form factors of the proton}",
    eprint = "2303.08347",
    archivePrefix = "arXiv",
    primaryClass = "hep-ph",
    reportNumber = "JLAB-PHY-23-3774",
    doi = "10.1103/RevModPhys.95.041002",
    journal = "Rev. Mod. Phys.",
    volume = "95",
    number = "4",
    pages = "041002",
    year = "2023"
}

@article{Polyakov:2018exb,
    author = "Polyakov, Maxim V. and Son, Hyeon-Dong",
    title = "{Nucleon gravitational form factors from instantons: forces between quark and gluon subsystems}",
    eprint = "1808.00155",
    archivePrefix = "arXiv",
    primaryClass = "hep-ph",
    doi = "10.1007/JHEP09(2018)156",
    journal = "JHEP",
    volume = "09",
    pages = "156",
    year = "2018"
}

@article{Hatta:2018sqd,
    author = "Hatta, Yoshitaka and Rajan, Abha and Tanaka, Kazuhiro",
    title = "{Quark and gluon contributions to the QCD trace anomaly}",
    eprint = "1810.05116",
    archivePrefix = "arXiv",
    primaryClass = "hep-ph",
    reportNumber = "YITP-18-112, J-PARC TH-0136, YITP-18-112,J-PARC TH-0136",
    doi = "10.1007/JHEP12(2018)008",
    journal = "JHEP",
    volume = "12",
    pages = "008",
    year = "2018"
}

@article{Tanaka:2022wzy,
    author = "Tanaka, Kazuhiro",
    title = "{Twist-four gravitational form factor at NNLO QCD from trace anomaly constraints}",
    eprint = "2212.09417",
    archivePrefix = "arXiv",
    primaryClass = "hep-ph",
    doi = "10.1007/JHEP03(2023)013",
    journal = "JHEP",
    volume = "03",
    pages = "013",
    year = "2023"
}

@article{CLAS:2015uuo,
    author = "Jo, H. S. and others",
    collaboration = "CLAS",
    title = "{Cross sections for the exclusive photon electroproduction on the proton and Generalized Parton Distributions}",
    eprint = "1504.02009",
    archivePrefix = "arXiv",
    primaryClass = "hep-ex",
    reportNumber = "JLAB-PHY-15-2037",
    doi = "10.1103/PhysRevLett.115.212003",
    journal = "Phys. Rev. Lett.",
    volume = "115",
    number = "21",
    pages = "212003",
    year = "2015"
}

@article{CLAS:2026lls,
    author = "Chatagnon, P. and others",
    collaboration = "CLAS",
    title = "{Measurement of the near-threshold J/{\ensuremath{\psi}} photoproduction cross section with the CLAS12 experiment}",
    eprint = "2602.22128",
    archivePrefix = "arXiv",
    primaryClass = "hep-ex",
    reportNumber = "JLAB-PHY-26-4605",
    doi = "10.1103/lqv1-kf2n",
    journal = "Phys. Rev. C",
    volume = "113",
    number = "6",
    pages = "065203",
    year = "2026"
}

@article{Kumericki:2016ehc,
    author = "Kumericki, Kresimir and Liuti, Simonetta and Moutarde, Herve",
    title = "{GPD phenomenology and DVCS fitting}: {Entering the high-precision era}",
    eprint = "1602.02763",
    archivePrefix = "arXiv",
    primaryClass = "hep-ph",
    doi = "10.1140/epja/i2016-16157-3",
    journal = "Eur. Phys. J. A",
    volume = "52",
    number = "6",
    pages = "157",
    year = "2016"
}

@article{Burkert:2018bqq,
    author = "Burkert, V. D. and Elouadrhiri, L. and Girod, F. X.",
    title = "{The pressure distribution inside the proton}",
    doi = "10.1038/s41586-018-0060-z",
    journal = "Nature",
    volume = "557",
    number = "7705",
    pages = "396--399",
    year = "2018"
}

@article{Burkert:2021ith,
    author = "Burkert, V. D. and Elouadrhiri, L. and Girod, F. X.",
    title = "{Determination of shear forces inside the proton}",
    eprint = "2104.02031",
    archivePrefix = "arXiv",
    primaryClass = "nucl-ex",
    month = "4",
    year = "2021"
}

@article{Kumericki:2019ddg,
    author = "Kumeri{\v{c}}ki, Kre{\v{s}}imir",
    title = "{Measurability of pressure inside the proton}",
    doi = "10.1038/s41586-019-1211-6",
    journal = "Nature",
    volume = "570",
    number = "7759",
    pages = "E1--E2",
    year = "2019"
}

@article{Dutrieux:2021nlz,
    author = "Dutrieux, H. and Lorc{\'e}, C. and Moutarde, H. and Sznajder, P. and Trawi{\'n}ski, A. and Wagner, J.",
    title = "{Phenomenological assessment of proton mechanical properties from deeply virtual Compton scattering}",
    eprint = "2101.03855",
    archivePrefix = "arXiv",
    primaryClass = "hep-ph",
    doi = "10.1140/epjc/s10052-021-09069-w",
    journal = "Eur. Phys. J. C",
    volume = "81",
    number = "4",
    pages = "300",
    year = "2021"
}

@article{Hashamipour:2022noy,
    author = "Hashamipour, Hadi and Goharipour, Muhammad and Azizi, K. and Goloskokov, S. V.",
    title = "{Generalized parton distributions at zero skewness}",
    eprint = "2211.09522",
    archivePrefix = "arXiv",
    primaryClass = "hep-ph",
    doi = "10.1103/PhysRevD.107.096005",
    journal = "Phys. Rev. D",
    volume = "107",
    number = "9",
    pages = "096005",
    year = "2023"
}

@article{Goharipour:2024atx,
    author = "Goharipour, Muhammad and Hashamipour, Hadi and Irani, Fatemeh and Azizi, K.",
    collaboration = "MMGPDs",
    title = "{Impact of JLab data on the determination of GPDs at zero skewness and
new insights from transition form factors $N\rightarrow \Delta$}",
    eprint = "2403.19384",
    archivePrefix = "arXiv",
    primaryClass = "hep-ph",
    doi = "10.1103/PhysRevD.109.074042",
    journal = "Phys. Rev. D",
    volume = "109",
    number = "7",
    pages = "074042",
    year = "2024"
}

@article{Selyugin:2023hqu,
    author = "Selyugin, O. V. and Teryaev, O. V.",
    title = "{GPDs and Gravitational Form Factors of Nucleons (Quark and Gluon Contributions)}",
    eprint = "2311.12988",
    archivePrefix = "arXiv",
    primaryClass = "hep-ph",
    doi = "10.1134/S1063778824700492",
    journal = "Phys. Atom. Nucl.",
    volume = "87",
    number = "4",
    pages = "518--522",
    year = "2024"
}

@article{GlueX:2023pev,
    author = "Adhikari, S. and others",
    collaboration = "GlueX",
    title = "{Measurement of the J/$\psi $ photoproduction cross section over the full near-threshold kinematic region}",
    eprint = "2304.03845",
    archivePrefix = "arXiv",
    primaryClass = "nucl-ex",
    reportNumber = "JLAB-PHY-23-3897",
    doi = "10.1103/PhysRevC.108.025201",
    journal = "Phys. Rev. C",
    volume = "108",
    number = "2",
    pages = "025201",
    year = "2023"
}

@article{Duran:2022xag,
    author = "Duran, B. and others",
    title = "{Determining the gluonic gravitational form factors of the proton}",
    eprint = "2207.05212",
    archivePrefix = "arXiv",
    primaryClass = "nucl-ex",
    doi = "10.1038/s41586-023-05730-4",
    journal = "Nature",
    volume = "615",
    number = "7954",
    pages = "813--816",
    year = "2023"
}

@article{Kharzeev:2021qkd,
    author = "Kharzeev, Dmitri E.",
    title = "{Mass radius of the proton}",
    eprint = "2102.00110",
    archivePrefix = "arXiv",
    primaryClass = "hep-ph",
    doi = "10.1103/PhysRevD.104.054015",
    journal = "Phys. Rev. D",
    volume = "104",
    number = "5",
    pages = "054015",
    year = "2021"
}

@article{Hatta:2018ina,
    author = "Hatta, Yoshitaka and Yang, Di-Lun",
    title = "{Holographic $J/\psi$ production near threshold and the proton mass problem}",
    eprint = "1808.02163",
    archivePrefix = "arXiv",
    primaryClass = "hep-ph",
    reportNumber = "YITP-18-79",
    doi = "10.1103/PhysRevD.98.074003",
    journal = "Phys. Rev. D",
    volume = "98",
    number = "7",
    pages = "074003",
    year = "2018"
}

@article{Mamo:2022eui,
    author = "Mamo, Kiminad A. and Zahed, Ismail",
    title = "{J/{\ensuremath{\psi}} near threshold in holographic QCD: A and D gravitational form factors}",
    eprint = "2204.08857",
    archivePrefix = "arXiv",
    primaryClass = "hep-ph",
    doi = "10.1103/PhysRevD.106.086004",
    journal = "Phys. Rev. D",
    volume = "106",
    number = "8",
    pages = "086004",
    year = "2022"
}

@article{Guo:2023pqw,
    author = "Guo, Yuxun and Ji, Xiangdong and Liu, Yizhuang and Yang, Jinghong",
    title = "{Updated analysis of near-threshold heavy quarkonium production for probe of proton{\textquoteright}s gluonic gravitational form factors}",
    eprint = "2305.06992",
    archivePrefix = "arXiv",
    primaryClass = "hep-ph",
    doi = "10.1103/PhysRevD.108.034003",
    journal = "Phys. Rev. D",
    volume = "108",
    number = "3",
    pages = "034003",
    year = "2023"
}

@article{Du:2020bqj,
    author = "Du, Meng-Lin and Baru, Vadim and Guo, Feng-Kun and Hanhart, Christoph and Mei{\ss}ner, Ulf-G and Nefediev, Alexey and Strakovsky, Igor",
    title = "{Deciphering the mechanism of near-threshold $J/\psi$ photoproduction}",
    eprint = "2009.08345",
    archivePrefix = "arXiv",
    primaryClass = "hep-ph",
    doi = "10.1140/epjc/s10052-020-08620-5",
    journal = "Eur. Phys. J. C",
    volume = "80",
    number = "11",
    pages = "1053",
    year = "2020"
}

@article{JointPhysicsAnalysisCenter:2023qgg,
    author = "Winney, D. and others",
    collaboration = "Joint Physics Analysis Center",
    title = "{Dynamics in near-threshold J/{\ensuremath{\psi}} photoproduction}",
    eprint = "2305.01449",
    archivePrefix = "arXiv",
    primaryClass = "hep-ph",
    reportNumber = "JLAB-THY-23-3802",
    doi = "10.1103/PhysRevD.108.054018",
    journal = "Phys. Rev. D",
    volume = "108",
    number = "5",
    pages = "054018",
    year = "2023"
}

@article{Tong:2021ctu,
    author = "Tong, Xuan-Bo and Ma, Jian-Ping and Yuan, Feng",
    title = "{Gluon gravitational form factors at large momentum transfer}",
    eprint = "2101.02395",
    archivePrefix = "arXiv",
    primaryClass = "hep-ph",
    doi = "10.1016/j.physletb.2021.136751",
    journal = "Phys. Lett. B",
    volume = "823",
    pages = "136751",
    year = "2021"
}

@article{Hatta:2021can,
    author = "Hatta, Yoshitaka and Strikman, Mark",
    title = "{$\phi$-meson lepto-production near threshold and the strangeness $D$-term}",
    eprint = "2102.12631",
    archivePrefix = "arXiv",
    primaryClass = "hep-ph",
    doi = "10.1016/j.physletb.2021.136295",
    journal = "Phys. Lett. B",
    volume = "817",
    pages = "136295",
    year = "2021"
}

@article{AbdulKhalek:2021gbh,
    author = "Abdul Khalek, R. and others",
    title = "{Science Requirements and Detector Concepts for the Electron-Ion Collider}: {EIC Yellow Report}",
    eprint = "2103.05419",
    archivePrefix = "arXiv",
    primaryClass = "physics.ins-det",
    reportNumber = "BNL-220990-2021-FORE, JLAB-PHY-21-3198, LA-UR-21-20953",
    doi = "10.1016/j.nuclphysa.2022.122447",
    journal = "Nucl. Phys. A",
    volume = "1026",
    pages = "122447",
    year = "2022"
}

@article{Accardi:2023chb,
    author = "Accardi, A. and others",
    title = "{Strong interaction physics at the luminosity frontier with 22 GeV electrons at Jefferson Lab}",
    eprint = "2306.09360",
    archivePrefix = "arXiv",
    primaryClass = "nucl-ex",
    reportNumber = "JLAB-PHY-23-3840, JLAB-THY-23-3848",
    doi = "10.1140/epja/s10050-024-01282-x",
    journal = "Eur. Phys. J. A",
    volume = "60",
    number = "9",
    pages = "173",
    year = "2024"
}

@article{Mathur:1999uf,
    author = "Mathur, N. and Dong, S. J. and Liu, K. F. and Mankiewicz, L. and Mukhopadhyay, N. C.",
    title = "{Quark orbital angular momentum from lattice QCD}",
    eprint = "hep-ph/9912289",
    archivePrefix = "arXiv",
    reportNumber = "SLAC-PUB-8331, UK-99-20",
    doi = "10.1103/PhysRevD.62.114504",
    journal = "Phys. Rev. D",
    volume = "62",
    pages = "114504",
    year = "2000"
}

@article{Hagler:2003jd,
    author = "Hagler, Philipp and Negele, John W. and Renner, Dru Bryant and Schroers, W. and Lippert, T. and Schilling, K.",
    collaboration = "LHPC, SESAM",
    title = "{Moments of nucleon generalized parton distributions in lattice QCD}",
    eprint = "hep-lat/0304018",
    archivePrefix = "arXiv",
    reportNumber = "MIT-CPT-3353",
    doi = "10.1103/PhysRevD.68.034505",
    journal = "Phys. Rev. D",
    volume = "68",
    pages = "034505",
    year = "2003"
}

@article{LHPC:2007blg,
    author = "Hagler, Ph. and others",
    collaboration = "LHPC",
    title = "{Nucleon Generalized Parton Distributions from Full Lattice QCD}",
    eprint = "0705.4295",
    archivePrefix = "arXiv",
    primaryClass = "hep-lat",
    reportNumber = "DESY-07-077, JLAB-THY-07-651, TUM-T39-07-09",
    doi = "10.1103/PhysRevD.77.094502",
    journal = "Phys. Rev. D",
    volume = "77",
    pages = "094502",
    year = "2008"
}

@article{Yang:2018bft,
    author = "Yang, Yi-Bo and Gong, Ming and Liang, Jian and Lin, Huey-Wen and Liu, Keh-Fei and Pefkou, Dimitra and Shanahan, Phiala",
    title = "{Nonperturbatively renormalized glue momentum fraction at the physical pion mass from lattice QCD}",
    eprint = "1805.00531",
    archivePrefix = "arXiv",
    primaryClass = "hep-lat",
    reportNumber = "JLAB-THY-18-2739",
    doi = "10.1103/PhysRevD.98.074506",
    journal = "Phys. Rev. D",
    volume = "98",
    number = "7",
    pages = "074506",
    year = "2018"
}

@article{Alexandrou:2020sml,
    author = "Alexandrou, C. and Bacchio, S. and Constantinou, M. and Finkenrath, J. and Hadjiyiannakou, K. and Jansen, K. and Koutsou, G. and Panagopoulos, H. and Spanoudes, G.",
    title = "{Complete flavor decomposition of the spin and momentum fraction of the proton using lattice QCD simulations at physical pion mass}",
    eprint = "2003.08486",
    archivePrefix = "arXiv",
    primaryClass = "hep-lat",
    doi = "10.1103/PhysRevD.101.094513",
    journal = "Phys. Rev. D",
    volume = "101",
    number = "9",
    pages = "094513",
    year = "2020"
}

@article{Liu:2021lke,
    author = "Liu, Keh-Fei",
    title = "{Status on lattice calculations of the proton spin decomposition}",
    eprint = "2112.08416",
    archivePrefix = "arXiv",
    primaryClass = "hep-lat",
    doi = "10.1007/s43673-022-00037-4",
    journal = "AAPPS Bull.",
    volume = "32",
    number = "1",
    pages = "8",
    year = "2022"
}

@article{Yang:2018nqn,
    author = "Yang, Yi-Bo and Liang, Jian and Bi, Yu-Jiang and Chen, Ying and Draper, Terrence and Liu, Keh-Fei and Liu, Zhaofeng",
    title = "{Proton Mass Decomposition from the QCD Energy Momentum Tensor}",
    eprint = "1808.08677",
    archivePrefix = "arXiv",
    primaryClass = "hep-lat",
    doi = "10.1103/PhysRevLett.121.212001",
    journal = "Phys. Rev. Lett.",
    volume = "121",
    number = "21",
    pages = "212001",
    year = "2018"
}

@article{Shanahan:2018pib,
    author = "Shanahan, P. E. and Detmold, W.",
    title = "{Gluon gravitational form factors of the nucleon and the pion from lattice QCD}",
    eprint = "1810.04626",
    archivePrefix = "arXiv",
    primaryClass = "hep-lat",
    reportNumber = "MIT-CTP/5069",
    doi = "10.1103/PhysRevD.99.014511",
    journal = "Phys. Rev. D",
    volume = "99",
    number = "1",
    pages = "014511",
    year = "2019"
}

@article{Pefkou:2021fni,
    author = "Pefkou, Dimitra A. and Hackett, Daniel C. and Shanahan, Phiala E.",
    title = "{Gluon gravitational structure of hadrons of different spin}",
    eprint = "2107.10368",
    archivePrefix = "arXiv",
    primaryClass = "hep-lat",
    reportNumber = "MIT-CTP/5318",
    doi = "10.1103/PhysRevD.105.054509",
    journal = "Phys. Rev. D",
    volume = "105",
    number = "5",
    pages = "054509",
    year = "2022"
}

@article{Shanahan:2018nnv,
    author = "Shanahan, P. E. and Detmold, W.",
    title = "{Pressure Distribution and Shear Forces inside the Proton}",
    eprint = "1810.07589",
    archivePrefix = "arXiv",
    primaryClass = "nucl-th",
    reportNumber = "MIT-CTP/5071",
    doi = "10.1103/PhysRevLett.122.072003",
    journal = "Phys. Rev. Lett.",
    volume = "122",
    number = "7",
    pages = "072003",
    year = "2019"
}

@article{Wang:2024lrm,
    author = "Wang, Bigeng and He, Fangcheng and Wang, Gen and Draper, Terrence and Liang, Jian and Liu, Keh-Fei and Yang, Yi-Bo",
    collaboration = "{\ensuremath{\chi}}QCD",
    title = "{Trace anomaly form factors from lattice QCD}",
    eprint = "2401.05496",
    archivePrefix = "arXiv",
    primaryClass = "hep-lat",
    doi = "10.1103/PhysRevD.109.094504",
    journal = "Phys. Rev. D",
    volume = "109",
    number = "9",
    pages = "094504",
    year = "2024"
}

@article{Hackett:2023rif,
    author = "Hackett, Daniel C. and Pefkou, Dimitra A. and Shanahan, Phiala E.",
    title = "{Gravitational Form Factors of the Proton from Lattice QCD}",
    eprint = "2310.08484",
    archivePrefix = "arXiv",
    primaryClass = "hep-lat",
    reportNumber = "MIT-CTP/5630, FERMILAB-PUB-23-592-T",
    doi = "10.1103/PhysRevLett.132.251904",
    journal = "Phys. Rev. Lett.",
    volume = "132",
    number = "25",
    pages = "251904",
    year = "2024"
}

@article{Belitsky:2002jp,
    author = "Belitsky, Andrei V. and Ji, X.",
    title = "{Chiral structure of nucleon gravitational form-factors}",
    eprint = "hep-ph/0203276",
    archivePrefix = "arXiv",
    reportNumber = "DOE-ER-40762-252, MD-PP-02-043",
    doi = "10.1016/S0370-2693(02)02025-7",
    journal = "Phys. Lett. B",
    volume = "538",
    pages = "289--297",
    year = "2002"
}

@article{Alharazin:2020yjv,
    author = "Alharazin, H. and Djukanovic, D. and Gegelia, J. and Polyakov, M. V.",
    title = "{Chiral theory of nucleons and pions in the presence of an external gravitational field}",
    eprint = "2006.05890",
    archivePrefix = "arXiv",
    primaryClass = "hep-ph",
    doi = "10.1103/PhysRevD.102.076023",
    journal = "Phys. Rev. D",
    volume = "102",
    number = "7",
    pages = "076023",
    year = "2020"
}

@article{Goeke:2007fp,
    author = "Goeke, K. and Grabis, J. and Ossmann, J. and Polyakov, M. V. and Schweitzer, P. and Silva, A. and Urbano, D.",
    title = "{Nucleon form-factors of the energy momentum tensor in the chiral quark-soliton model}",
    eprint = "hep-ph/0702030",
    archivePrefix = "arXiv",
    reportNumber = "RUB-TP2-05-2006",
    doi = "10.1103/PhysRevD.75.094021",
    journal = "Phys. Rev. D",
    volume = "75",
    pages = "094021",
    year = "2007"
}

@article{Won:2023zmf,
    author = "Won, Ho-Yeon and Kim, Hyun-Chul and Kim, June-Young",
    title = "{Mechanical structure of the nucleon and the baryon octet: twist-2 case}",
    eprint = "2310.04670",
    archivePrefix = "arXiv",
    primaryClass = "hep-ph",
    reportNumber = "JLAB-THY-23-3938",
    doi = "10.1007/JHEP05(2024)173",
    journal = "JHEP",
    volume = "05",
    pages = "173",
    year = "2024"
}

@article{Cebulla:2007ei,
    author = "Cebulla, C. and Goeke, K. and Ossmann, J. and Schweitzer, P.",
    title = "{The Nucleon form-factors of the energy momentum tensor in the Skyrme model}",
    eprint = "hep-ph/0703025",
    archivePrefix = "arXiv",
    reportNumber = "RUB-TP2-01-2007",
    doi = "10.1016/j.nuclphysa.2007.08.004",
    journal = "Nucl. Phys. A",
    volume = "794",
    pages = "87--114",
    year = "2007"
}

@article{Tanaka:2025pny,
    author = "Tanaka, Mitsuru and Fujii, Daisuke and Kawaguchi, Mamiya",
    title = "{Gravitational form factors of the nucleon in the Skyrme model based on scale-invariant chiral perturbation theory}",
    eprint = "2507.21220",
    archivePrefix = "arXiv",
    primaryClass = "hep-ph",
    doi = "10.1103/gzj5-7bln",
    journal = "Phys. Rev. D",
    volume = "112",
    number = "5",
    pages = "054048",
    year = "2025"
}

@article{Neubelt:2019sou,
    author = "Neubelt, Matt J. and Sampino, Andrew and Hudson, Jonathan and Tezgin, Kemal and Schweitzer, Peter",
    title = "{Energy momentum tensor and the D-term in the bag model}",
    eprint = "1911.08906",
    archivePrefix = "arXiv",
    primaryClass = "hep-ph",
    doi = "10.1103/PhysRevD.101.034013",
    journal = "Phys. Rev. D",
    volume = "101",
    number = "3",
    pages = "034013",
    year = "2020"
}

@article{Pasquini:2014vua,
    author = "Pasquini, B. and Polyakov, M. V. and Vanderhaeghen, M.",
    title = "{Dispersive evaluation of the D-term form factor in deeply virtual Compton scattering}",
    eprint = "1407.5960",
    archivePrefix = "arXiv",
    primaryClass = "hep-ph",
    doi = "10.1016/j.physletb.2014.10.047",
    journal = "Phys. Lett. B",
    volume = "739",
    pages = "133--138",
    year = "2014"
}

@article{Cao:2024zlf,
    author = "Cao, Xiong-Hui and Guo, Feng-Kun and Li, Qu-Zhi and Yao, De-Liang",
    title = "{Dispersive determination of nucleon gravitational form factors}",
    eprint = "2411.13398",
    archivePrefix = "arXiv",
    primaryClass = "hep-ph",
    doi = "10.1038/s41467-025-62278-9",
    journal = "Nature Commun.",
    volume = "16",
    pages = "6979",
    year = "2025"
}

@article{Abidin:2009hr,
    author = "Abidin, Zainul and Carlson, Carl E.",
    title = "{Nucleon electromagnetic and gravitational form factors from holography}",
    eprint = "0903.4818",
    archivePrefix = "arXiv",
    primaryClass = "hep-ph",
    doi = "10.1103/PhysRevD.79.115003",
    journal = "Phys. Rev. D",
    volume = "79",
    pages = "115003",
    year = "2009"
}

@article{Chakrabarti:2020kdc,
    author = "Chakrabarti, Dipankar and Mondal, Chandan and Mukherjee, Asmita and Nair, Sreeraj and Zhao, Xingbo",
    title = "{Gravitational form factors and mechanical properties of proton in a light-front quark-diquark model}",
    eprint = "2010.04215",
    archivePrefix = "arXiv",
    primaryClass = "hep-ph",
    doi = "10.1103/PhysRevD.102.113011",
    journal = "Phys. Rev. D",
    volume = "102",
    pages = "113011",
    year = "2020"
}

@article{Choudhary:2022den,
    author = "Choudhary, Poonam and Gurjar, Bheemsehan and Chakrabarti, Dipankar and Mukherjee, Asmita",
    title = "{Gravitational form factors and mechanical properties of the proton: Connections between distributions in 2D and 3D}",
    eprint = "2206.12206",
    archivePrefix = "arXiv",
    primaryClass = "hep-ph",
    doi = "10.1103/PhysRevD.106.076004",
    journal = "Phys. Rev. D",
    volume = "106",
    number = "7",
    pages = "076004",
    year = "2022"
}

@article{Nair:2025sfr,
    author = "Nair, Sreeraj and Mondal, Chandan and Xu, Siqi and Zhao, Xingbo and Vary, James P.",
    collaboration = "BLFQ",
    title = "{Proton gravitational structure and mass decomposition on the light front}",
    eprint = "2506.07554",
    archivePrefix = "arXiv",
    primaryClass = "hep-ph",
    doi = "10.1103/79fs-5ghd",
    journal = "Phys. Rev. D",
    volume = "112",
    number = "11",
    pages = "114001",
    year = "2025"
}

@article{Freese:2020mcx,
    author = {Freese, Adam and Clo{\"e}t, Ian C.},
    title = "{Quark spin and orbital angular momentum from proton generalized parton distributions}",
    eprint = "2005.10286",
    archivePrefix = "arXiv",
    primaryClass = "nucl-th",
    doi = "10.1103/PhysRevC.103.045204",
    journal = "Phys. Rev. C",
    volume = "103",
    number = "4",
    pages = "045204",
    year = "2021"
}

@article{Qiu:2026kde,
    author = "Qiu, Yiwei and Liu, Zhibo and Zhang, Congcong",
    title = "{Gravitational form factors of the baryon octet and flavor dependence in an NJL quark--diquark model}",
    eprint = "2611.01948",
    archivePrefix = "arXiv",
    primaryClass = "hep-ph",
    month = "11",
    year = "2026"
}

@article{Yao:2024ixu,
    author = "Yao, Z. -Q. and Xu, Y. -Z. and Binosi, D. and Cui, Z. -F. and Ding, M. and Raya, K. and Roberts, C. D. and Rodr{\'\i}guez-Quintero, J. and Schmidt, S. M.",
    title = "{Nucleon gravitational form factors}",
    eprint = "2409.15547",
    archivePrefix = "arXiv",
    primaryClass = "hep-ph",
    reportNumber = "NJU-INP 091/24",
    doi = "10.1140/epja/s10050-025-01557-x",
    journal = "Eur. Phys. J. A",
    volume = "61",
    number = "5",
    pages = "92",
    year = "2025"
}

@article{Broniowski:2025ctl,
    author = "Broniowski, Wojciech and Ruiz Arriola, Enrique",
    title = "{Gravitational form factors and mechanical properties of the nucleon in a meson dominance approach}",
    eprint = "2503.09297",
    archivePrefix = "arXiv",
    primaryClass = "hep-ph",
    doi = "10.1103/ylml-mrlh",
    journal = "Phys. Rev. D",
    volume = "112",
    number = "5",
    pages = "054028",
    year = "2025"
}

@article{Varma:2020crx,
    author = "Varma, Mira and Schweitzer, Peter",
    title = "{Effects of long-range forces on the D-term and the energy-momentum structure}",
    eprint = "2006.06602",
    archivePrefix = "arXiv",
    primaryClass = "hep-ph",
    doi = "10.1103/PhysRevD.102.014047",
    journal = "Phys. Rev. D",
    volume = "102",
    number = "1",
    pages = "014047",
    year = "2020"
}

@article{Metz:2021lqv,
    author = "Metz, Andreas and Pasquini, Barbara and Rodini, Simone",
    title = "{The gravitational form factor D(t) of the electron}",
    eprint = "2104.04207",
    archivePrefix = "arXiv",
    primaryClass = "hep-ph",
    doi = "10.1016/j.physletb.2021.136501",
    journal = "Phys. Lett. B",
    volume = "820",
    pages = "136501",
    year = "2021"
}

@article{Hackett:2023nkr,
    author = "Hackett, Daniel C. and Oare, Patrick R. and Pefkou, Dimitra A. and Shanahan, Phiala E.",
    title = "{Gravitational form factors of the pion from lattice QCD}",
    eprint = "2307.11707",
    archivePrefix = "arXiv",
    primaryClass = "hep-lat",
    reportNumber = "MIT-CTP/5585",
    doi = "10.1103/PhysRevD.108.114504",
    journal = "Phys. Rev. D",
    volume = "108",
    number = "11",
    pages = "114504",
    year = "2023"
}

@article{Kim:2020lrs,
    author = "Kim, June-Young and Sun, Bao-Dong",
    title = "{Gravitational form factors of a baryon with spin-3/2}",
    eprint = "2011.00292",
    archivePrefix = "arXiv",
    primaryClass = "hep-ph",
    doi = "10.1140/epjc/s10052-021-08852-z",
    journal = "Eur. Phys. J. C",
    volume = "81",
    number = "1",
    pages = "85",
    year = "2021"
}

@article{Dehghan:2023ytx,
    author = {Dehghan, Z. and Azizi, K. and {\"O}zdem, U.},
    title = "{Gravitational form factors of the {\ensuremath{\Delta}} baryon via QCD sum rules}",
    eprint = "2307.14880",
    archivePrefix = "arXiv",
    primaryClass = "hep-ph",
    doi = "10.1103/PhysRevD.108.094037",
    journal = "Phys. Rev. D",
    volume = "108",
    number = "9",
    pages = "094037",
    year = "2023"
}

@article{Dehghan:2025eov,
    author = "Dehghan, Zeinab and Azizi, K.",
    title = "{Mechanical properties of the {\ensuremath{\Omega}}- baryon from gravitational form factors}",
    eprint = "2507.14840",
    archivePrefix = "arXiv",
    primaryClass = "hep-ph",
    doi = "10.1103/x6f5-cbcc",
    journal = "Phys. Rev. D",
    volume = "112",
    number = "5",
    pages = "054014",
    year = "2025"
}

@article{GarciaMartin-Caro:2023klo,
    author = "Garcia Martin-Caro, Alberto and Huidobro, Miguel and Hatta, Yoshitaka",
    title = "{Gravitational form factors of nuclei in the Skyrme model}",
    eprint = "2304.05994",
    archivePrefix = "arXiv",
    primaryClass = "nucl-th",
    doi = "10.1103/PhysRevD.108.034014",
    journal = "Phys. Rev. D",
    volume = "108",
    number = "3",
    pages = "034014",
    year = "2023"
}

@article{Mai:2012yc,
    author = "Mai, Manuel and Schweitzer, Peter",
    title = "{Energy momentum tensor, stability, and the D-term of Q-balls}",
    eprint = "1206.2632",
    archivePrefix = "arXiv",
    primaryClass = "hep-ph",
    doi = "10.1103/PhysRevD.86.076001",
    journal = "Phys. Rev. D",
    volume = "86",
    pages = "076001",
    year = "2012"
}

@article{Shifman:1978bx,
    author = "Shifman, Mikhail A. and Vainshtein, A. I. and Zakharov, Valentin I.",
    title = "{QCD and Resonance Physics. Theoretical Foundations}",
    reportNumber = "ITEP-73-1978, ITEP-80-1978",
    doi = "10.1016/0550-3213(79)90022-1",
    journal = "Nucl. Phys. B",
    volume = "147",
    pages = "385--447",
    year = "1979"
}

@article{Ioffe:1981kw,
    author = "Ioffe, B. L.",
    title = "{Calculation of Baryon Masses in Quantum Chromodynamics}",
    reportNumber = "ITEP-10-1981, ITEP-92-1981",
    doi = "10.1016/0550-3213(81)90259-5",
    journal = "Nucl. Phys. B",
    volume = "188",
    pages = "317--341",
    year = "1981",
    note = "[Erratum: Nucl.Phys.B 191, 591--592 (1981)]"
}

@article{Anikin:2019kwi,
    author = "Anikin, I. V.",
    title = "{Gravitational form factors within light-cone sum rules at leading order}",
    eprint = "1902.00094",
    archivePrefix = "arXiv",
    primaryClass = "hep-ph",
    doi = "10.1103/PhysRevD.99.094026",
    journal = "Phys. Rev. D",
    volume = "99",
    number = "9",
    pages = "094026",
    year = "2019"
}

@article{Azizi:2019ytx,
    author = {Azizi, K. and {\"O}zdem, U.},
    title = "{Nucleon{\textquoteright}s energy{\textendash}momentum tensor form factors in light-cone QCD}",
    eprint = "1908.06143",
    archivePrefix = "arXiv",
    primaryClass = "hep-ph",
    doi = "10.1140/epjc/s10052-020-7676-5",
    journal = "Eur. Phys. J. C",
    volume = "80",
    number = "2",
    pages = "104",
    year = "2020"
}

@article{Dehghan:2025ncw,
    author = "Dehghan, Zeinab and Almaksusi, F. and Azizi, K.",
    title = "{Mechanical properties of proton using flavor-decomposed gravitational form factors}",
    eprint = "2502.16689",
    archivePrefix = "arXiv",
    primaryClass = "hep-ph",
    doi = "10.1007/JHEP06(2025)025",
    journal = "JHEP",
    volume = "06",
    pages = "025",
    year = "2025"
}

@article{Ozdem:2020ieh,
    author = {{\"O}zdem, U. and Azizi, K.},
    title = "{Gravitational form factors of hyperons in light-cone QCD}",
    eprint = "2003.12588",
    archivePrefix = "arXiv",
    primaryClass = "hep-ph",
    doi = "10.1103/PhysRevD.101.114026",
    journal = "Phys. Rev. D",
    volume = "101",
    number = "11",
    pages = "114026",
    year = "2020"
}

@article{Azizi:2020jog,
    author = {Azizi, K. and {\"O}zdem, U.},
    title = "{Gravitational form factors of N(1535) in QCD}",
    eprint = "2012.06895",
    archivePrefix = "arXiv",
    primaryClass = "hep-ph",
    doi = "10.1016/j.nuclphysa.2021.122296",
    journal = "Nucl. Phys. A",
    volume = "1015",
    pages = "122296",
    year = "2021"
}

@article{Ozdem:2019pkg,
    author = {{\"O}zdem, U. and Azizi, K.},
    title = "{Gravitational transition form factors of $N(1535) \rightarrow N$}",
    eprint = "1912.06375",
    archivePrefix = "arXiv",
    primaryClass = "hep-ph",
    doi = "10.1103/PhysRevD.101.054031",
    journal = "Phys. Rev. D",
    volume = "101",
    number = "5",
    pages = "054031",
    year = "2020"
}

@article{Ozdem:2022zig,
    author = {{\"O}zdem, U. and Azizi, K.},
    title = "{Gravitational transition form factors of N {\textrightarrow} {\ensuremath{\Delta}} via QCD light-cone sum rules}",
    eprint = "2212.07290",
    archivePrefix = "arXiv",
    primaryClass = "hep-ph",
    doi = "10.1007/JHEP03(2023)048",
    journal = "JHEP",
    volume = "03",
    pages = "048",
    year = "2023"
}

@article{Asmaee:2026nhr,
    author = "Asmaee, Z. and Azizi, K.",
    title = "{Photon and gluon gravitational form factors of proton in QCD}",
    eprint = "2608.23788",
    archivePrefix = "arXiv",
    primaryClass = "hep-ph",
    month = "8",
    year = "2026"
}

@article{Ioffe:1982qb,
    author = "Ioffe, B. L. and Smilga, Andrei V.",
    title = "{Meson Widths and Form-Factors at Intermediate Momentum Transfer in Nonperturbative QCD}",
    reportNumber = "ITEP-100-1982",
    doi = "10.1016/0550-3213(83)90291-2",
    journal = "Nucl. Phys. B",
    volume = "216",
    pages = "373--407",
    year = "1983"
}

@article{Nesterenko:1982gc,
    author = "Nesterenko, V. A. and Radyushkin, A. V.",
    title = "{Sum Rules and Pion Form-Factor in QCD}",
    reportNumber = "JINR-E2-82-204",
    doi = "10.1016/0370-2693(82)90528-7",
    journal = "Phys. Lett. B",
    volume = "115",
    pages = "410",
    year = "1982"
}

@article{ParticleDataGroup:2024cfk,
    author = "Navas, S. and others",
    collaboration = "Particle Data Group",
    title = "{Review of particle physics}",
    doi = "10.1103/PhysRevD.110.030001",
    journal = "Phys. Rev. D",
    volume = "110",
    number = "3",
    pages = "030001",
    year = "2024"
}

@article{Belyaev:1982sa,
    author = "Belyaev, V. M. and Ioffe, B. L.",
    title = "{Determination of Baryon and Baryonic Resonance Masses from QCD Sum Rules. 1. Nonstrange Baryons}",
    reportNumber = "ITEP-59-1982",
    journal = "Sov. Phys. JETP",
    volume = "56",
    pages = "493--501",
    year = "1982"
}

@article{Ioffe:2005ym,
    author = "Ioffe, B. L.",
    title = "{QCD at low energies}",
    eprint = "hep-ph/0502148",
    archivePrefix = "arXiv",
    doi = "10.1016/j.ppnp.2005.05.001",
    journal = "Prog. Part. Nucl. Phys.",
    volume = "56",
    pages = "232--277",
    year = "2006"
}

@article{Bakulev:1991ps,
    author = "Bakulev, A. P. and Radyushkin, A. V.",
    title = "{Nonlocal condensates and QCD sum rules for the pion form factor}",
    reportNumber = "CEBAF-TH-91-10",
    doi = "10.1016/0370-2693(91)91304-E",
    journal = "Phys. Lett. B",
    volume = "271",
    pages = "223--230",
    year = "1991"
}

@article{Bakulev:2009ib,
    author = "Bakulev, A. P. and Pimikov, A. V. and Stefanis, N. G.",
    title = "{QCD sum rules with nonlocal condensates and the spacelike pion form factor}",
    eprint = "0904.2304",
    archivePrefix = "arXiv",
    primaryClass = "hep-ph",
    reportNumber = "RUB-TPII-03-09",
    doi = "10.1103/PhysRevD.79.093010",
    journal = "Phys. Rev. D",
    volume = "79",
    pages = "093010",
    year = "2009"
}

@article{Mikhailov:1991pt,
    author = "Mikhailov, S. V. and Radyushkin, A. V.",
    title = "{The Pion wave function and QCD sum rules with nonlocal condensates}",
    reportNumber = "CEBAF-TH-91-11",
    doi = "10.1103/PhysRevD.45.1754",
    journal = "Phys. Rev. D",
    volume = "45",
    pages = "1754--1759",
    year = "1992"
}

\end{document}